# Cell-cell communication enhances the capacity of cell ensembles to sense shallow gradients during morphogenesis

David Ellison[1,4§], Andrew Mugler[2,3§], Matthew Brennan[1,4§], Sung Hoon Lee[4], Robert Huebner[5], Eliah Shamir[5], Laura A. Woo[1], Joseph Kim[1], Patrick Amar[6], Ilya Nemenman[2,7*], Andrew J. Ewald[5*], Andre Levchenko[4*]

1. Department of Biomedical Engineering, Johns Hopkins University, Baltimore, MD 21218, USA
2. Department of Physics, Emory University, Atlanta, GA 30322, USA
3. Department of Physics and Astronomy, Purdue University, West Lafayette, IN 47907, USA
4. Department of Biomedical Engineering and Yale Systems Biology Institute, Yale University, New Haven, CT 06520, USA
5. Center for Cell Dynamics, Johns Hopkins University, Baltimore, MD 21205, USA
6. Université Paris-Sud, 91405 ORSAY Cedex, France
7. Department of Biology, Emory University, Atlanta, GA 30322, USA

§ These authors contributed equally to this work
* Corresponding authors: ilya.nemenman@emory.edu, andrew.ewald@jhmi.edu, and andre.levchenko@yale.edu

## Abstract

Collective cell responses to exogenous cues depend on cell-cell interactions. In principle, these can result in enhanced sensitivity to weak and noisy stimuli. However, this has not yet been shown experimentally, and, little is known about how multicellular signal processing modulates single cell sensitivity to extracellular signaling inputs, including those guiding complex changes in the tissue form and function. Here we explored if cell-cell communication can enhance the ability of cell ensembles to sense and respond to weak gradients of chemotactic cues. Using a combination of experiments with mammary epithelial cells and mathematical modeling, we find that multicellular sensing enables detection of and response to shallow Epidermal Growth Factor (EGF) gradients that are undetectable by single cells. However, the advantage of this type of gradient sensing is limited by the noisiness of the signaling relay, necessary to integrate spatially distributed ligand concentration information. We calculate the fundamental sensory limits imposed by this communication noise and combine them with the experimental data to estimate the effective size of multicellular sensory groups involved in gradient sensing. Functional experiments strongly implicated intercellular communication through gap junctions and calcium release from intracellular stores as mediators of collective gradient sensing. The resulting integrative analysis provides a framework for understanding the advantages and limitations of sensory information processing by relays of chemically coupled cells.

## Significance Statement

What new properties may result from collective cell behavior, and how these emerging capabilities may influence shaping and function of tissues, in health and disease? Here, we explored these questions in the context of epithelial branching morphogenesis. We show experimentally that, while individual mammary epithelial cells are incapable of sensing extremely weak gradients of a growth factor, cellular collectives in organotypic cultures exhibit reliable, gradient driven, directional growth. This underscores a critical importance of collective cell-cell communication and computation in gradient sensing. We develop and verify a biophysical theory of such communication, and identify the mechanisms by which it is implemented in the mammary epithelium, quantitatively analyzing both advantages and limitations of biochemical cellular communication in collective decision making.

## Introduction

Responses of isogenic cells to identical cues can display considerable variability. For instance, a population of cells will typically exhibit substantial variation in gradient sensitivity and migration trajectories within the same gradient of a diffusible guidance signal (1). The variation in response could arise from the inherent diversity of cell responsiveness (2-5), but it can be further exacerbated if the gradients of extracellular signals are shallow and noisy (6-11). In fact, sensing shallow gradients can approach fundamental physical limits that define whether diffusive graded cues can bias cell migration (12, 13). However, the spatially biased response can improve and its uncertainty can be substantially reduced if individual cells are coupled while responding to molecular gradients (5, 14-21). Strong cell-cell coupling might reduce the response noise by averaging individual responses of multiple cells (22-27). It can also alleviate sensory noise by extending the spatial range of the sensing, thus increasing the potential for more precise detection of weak and noisy spatially graded inputs. Importantly, however, cell-cell communication involved in such collective sensing may be itself subject to noise, reducing the precision of the communicated signals and therefore the advantage gained from an augmented size of the sensory and the response units. The interplay between the increasing signal and accumulating communication noise associated with the multicellular sensing, and thus the limits of this multicellular sensing strategy, remain incompletely understood.

An example of collective cellular response is branching morphogenesis of the epithelial tissue in mammary glands (28-30). The dynamic processes, whose coordinate regulation leads to formation, growth, and overall organization of branched epithelial structures, are still actively investigated (29). Conveniently, the morphogenesis of mammary glands is recapitulated in organotypic mammary culture (organoids) (31-33), extensively used to model and explore various features of self-organization and development of epithelial tissues (34). Epidermal Growth Factor (EGF) is an essential regulator of branching morphogenesis in mammary glands (35, 36). It has also been identified as a critical chemo-attractant guiding the migration of breast epithelial cells in invasive cancer growth (37). This property of EGF raises the possibility that it can serve as an endogenous chemo-attractant guiding formation and extension of mammary epithelial branches, a possibility that has not yet been experimentally addressed.

Our data reveal that the capacity of mammary organoids embedded in collagen I to respond to shallow EGF gradients requires collective gradient sensing, mediated by intercellular chemical coupling though gap junctions. Surprisingly, the advantage of multicellular sensing is limited and is substantially lower than the theoretical predictions stemming from gradient sensing models that do not account for communication noise (6). We build a theory of the multicellular sensing process, equivalent to the information-theoretic relay channel, which correctly predicts the accuracy of sensing as a function of the gradient magnitude, organoid size, and the background ligand concentration. The theory and the corresponding stochastic computational model trace the reduced sensing improvement to the unavoidable noise in the information relay used by cells to transmit their local sensory measurements to each other. This analysis allowed us to determine the approximate size of a collective, multicellular sensing unit enabling chemotropic branch formation and growth.

## Results

To study the response of multicellular mammary organoids to defined growth factor gradients, we developed and used mesoscopic fluidic devices. These devices permitted generation of highly controlled gradients of EGF, that were stable for a few days, within small slabs of collagen gels housing expanding organoids (see Fig. 1A, Methods, and Supplementary Information). We found that organoids of diverse sizes, ranging from 80 to 300 µm (or about 200 to 500 cells), developed normally within the device, forming multiple branches in the presence of spatially uniform 2.5 nM of EGF. When monitored over 3 days, the branch formation in such uniformly distributed EGF displayed no directional bias (Fig. 1C; Supplementary Fig. S2A). However, if EGF was added as a very shallow gradient of $0.5 \times 10^{-3}$ nM/µm (equivalent to about $0.5 \times 10^{-2}$ nM or as little as 0.2% concentration difference across a 10 µm cell), branch formation displayed a significant directional bias (Fig. 1D). The bias in formation of new branches remained the same when measured on each of the three consecutive days (Supp. Fig. S4), suggesting that EGF gradient sensing is not a transient response, and that its angular precision neither improves nor decreases with time. The bias was robust to the choice of the bias measure, as six different measures all yielded values at least four standard errors above their respective null values (Fig. 2; see also *SI* and Supp. Fig. S1).

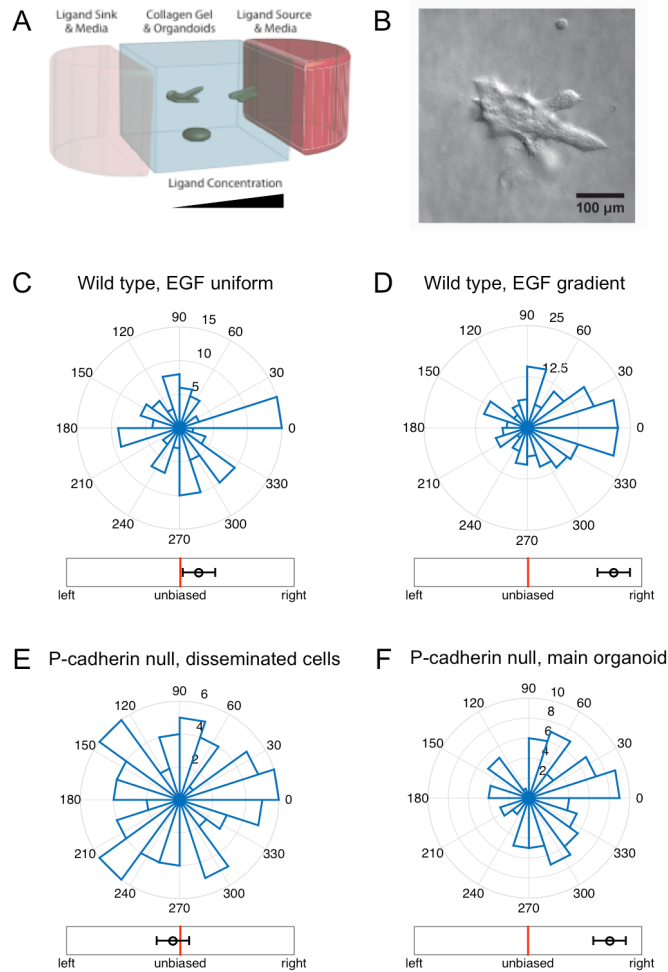

**Figure 1. Organoid branching but not single cell migration exhibits biased directional response to an EGF ligand gradient.** (A) A schematic of the mesofluidic device chamber, with high (red) and low (pink) EGF concentration reservoirs and organoids embedded in collagen gel exposed to the resulting EGF gradient (See Supplementary Materials for further information). (B) Example microscopy image of an organoid exposed to a 0.5 nM/mm EGF gradient, with preferential branch formation in the gradient direction (toward the right side of the field of view). (C-F) Angular histograms of (C) organoid branching directions in a uniform background of 2.5 nM EGF (3 biological replicates, 8 experimental replicates, total 110 organoids, total 460 branches); (D) organoid branching directions in a gradient of 0.5 nM/mm EGF (2 biological replicates, 2 experimental replicates, total 200 organoids, total 1283 branches); (E) in-gel migration directions of single cells separating from P-cadherin deficient organoids in a gradient of 0.5 nM/mm EGF (2 biological replicates, 6 experimental replicates, total 255 cells originating from 76 organoids); (F) organoid branching directions for the P- cadherin organoids in a gradient of 0.5 nM/mm EGF (2 biological replicates, 6 experimental replicates, total 79 organoids, total 394 branches). Whereas single cells do not exhibit biased movement, organoids exhibit biased branching. In (C), (D) and (F), branching direction is defined as the angle of the vector sum of the organoid's branches. The number of organoids branching in a specific direction is shown. In (E), cell migration direction is defined as the angle of the vector sum of the displacements of all cells originating from a given organoid. In (C-F), the circular axes measure the number of organoids with that branching direction, and left-right bias underneath each histogram is defined in Fig. 2 (measure B).

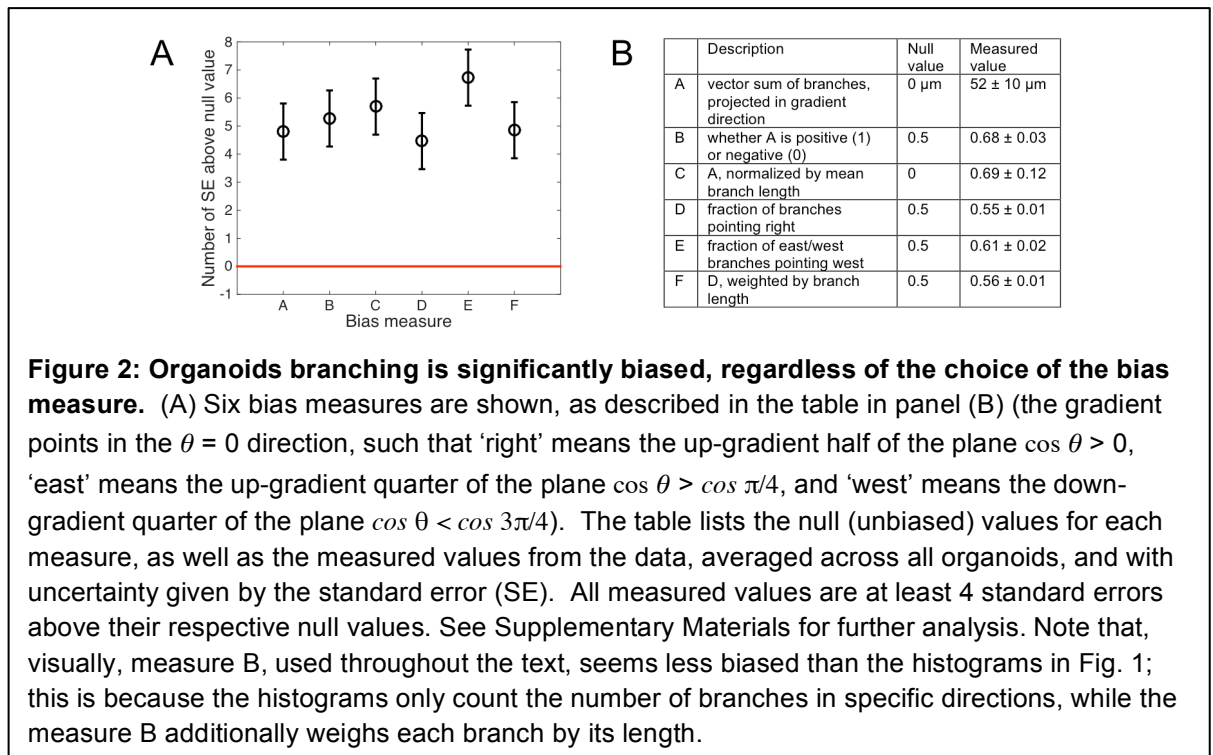

**Figure 2: Organoids branching is significantly biased, regardless of the choice of the bias measure.** (A) Six bias measures are shown, as described in the table in panel (B) (the gradient points in the $\theta = 0$ direction, such that 'right' means the up-gradient half of the plane $\cos \theta > 0$, 'east' means the up-gradient quarter of the plane $\cos \theta > \cos \pi/4$, and 'west' means the down-gradient quarter of the plane $\cos \theta < \cos 3\pi/4$). The table lists the null (unbiased) values for each measure, as well as the measured values from the data, averaged across all organoids, and with uncertainty given by the standard error (SE). All measured values are at least 4 standard errors above their respective null values. See Supplementary Materials for further analysis. Note that, visually, measure B, used throughout the text, seems less biased than the histograms in Fig. 1; this is because the histograms only count the number of branches in specific directions, while the measure B additionally weighs each branch by its length.

In spite of the very shallow EGF gradient, it was still possible that the spatial bias in branching was a consequence of the gradient sensing by individual cells within the tips of the branches. To examine the sensitivity of single cells to these shallow gradients in the 3D geometry of collagen gels, we analyzed organoids derived from P-Cadherin knock out mice (38). Consistent with our previous findings (39), the luminal epithelial core of the organoids derived from P-cadherin null mice remain intact within collagen I gels, but individual and small groups of myoepithelial cells disseminate into the surrounding gel, since P-cadherin is a specific mediator of myoepithelial cell-cell adhesion. These individual dissociated cells displayed extensive migration through the collagen matrix. Although in these experiments the organoids continued to display EGF gradient-guided directional branching responses similar to those of WT organoids, the dissociated cells migrated in a completely unbiased manner (Fig. 1E, F; Supp. Figs. S2B, S3; see also *SI*). Cell motility and the distance traveled by single cells within the gels generally were the same as those observed in similar experiments performed in spatially homogenous 2.5 nM EGF distributions (data not shown). These results were corroborated by experiments in which dissociated single mammary epithelial cells isolated from WT mice or MTLn3-B1 cells were embedded in the same devices and subjected to the same experimental inputs (Supp. Fig. S4). The results of these experiments suggested that, in spite of considerable motility, there was no evidence of chemotaxis by these cells, in response to EGF gradients that were capable of triggering biased chemotropic response in organoids. Overall, our results reveal that cell-cell coupling within organoids permits sensing of EGF gradients not detectable by single cells.

Can enhanced collective gradient sensing by multiple cells be explained by a quantitative theory,

permitting experimental validation? The classic Berg-Purcell (BP) theory of concentration (40) and gradient (6, 12) detection can explain why a larger detector (in this case, an organoid) has a better sensitivity than a smaller one (a cell). Briefly, the mean number of ligand molecules in the volume of a detector of a linear size $A$ is $\bar{\nu} \sim \bar{c}A^3$, where $c$ is the concentration being determined, and overbar represents averaging. This number is Poisson distributed, so that the relative error in counting is $(\delta\nu/\bar{\nu})^2 = (\delta c/\bar{c})^2 = 1/\bar{\nu} \sim 1/(\bar{c}A^3)$. This bound can be modified to include temporal integration of the ligand diffusing in and out of the receptor vicinity (41). However, the organoids show steady branching and no improved directional sensitivity over the three days of experiments (Supp. Fig. S5), so the sensing can be assumed to occur on time scales much shorter than the overall branching response, without long-time integration. Estimation of spatial gradients by a cell or a multicellular ensemble involves inference of the difference between (or comparison of) concentrations measured by different compartments of the detector (6, 8, 12) (branches grow too slowly for a temporal comparison strategy to be useful (42)). For a detector consisting of two such compartments, each of size $a \ll A$, the mean concentration in each compartment is $\bar{c}_\pm = \bar{c}_{1/2} \pm gA/2$, where $c_{1/2}$ is the concentration at the center of the detector, and $g$ is the concentration gradient. For each of the compartments, the BP bound gives $(\delta c_\pm/\bar{c}_\pm)^2 \sim 1/(\bar{c}_\pm a^3)$. Subtracting the two independently measured concentrations estimates the gradient, $g = (c_+ - c_-)/A$, which results in the signal-to-noise ratio (SNR, or inverse of the error):

$$\frac{1}{\mathrm{SNR}} \equiv \left(\frac{\delta g}{g}\right)^2 \approx \frac{\bar{c}_{1/2}}{a^3 (Ag)^2}. \quad [1]$$

Thus the sensing precision should improve without bound with the span of the gradient being measured ($A$), with the gradient strength ($g$), and with the volume over which molecules are counted ($a^3$). However, the precision should decrease with the background concentration ($\bar{c}_{1/2}$) because it is hard to measure small changes in a signaling molecule against a large background concentration of this molecule.[1] Note that Eq. [1] seems to predict an infinitely precise measurement when $\bar{c}_{1/2} \to 0$, and there are no ligand molecules. This paradox is resolved by the simple observation that the background concentration of the signaling molecule and the organoid size are not independent: in a linear gradient, $\bar{c}_{1/2}$ is limited from below by $Ag/2$, and, generally, small $\bar{c}_{1/2}$ is only possible for a small organoid if the gradient is nonzero. In this *low concentration* limit of the BP theory, which is often the subject of analysis (6, 12), $Ag \sim \bar{c}_{1/2}$. Then Eq. [1] transforms to $\mathrm{SNR} \sim \bar{c}_{1/2} a^3$, and the SNR *increases* with $\bar{c}_{1/2}$. Overall, this interplay between the size and the concentration depends on $\bar{c}_{1/2}(A)$, which may take different forms depending on where organoids of different sizes are in the experimental device. Typically, the SNR has an inverted U-shape: it first grows with $\bar{c}_{1/2}$ because the span of the organoid increases, and then it drops because small differences of large concentrations must be estimated by a cell or a cell ensemble (see *SI* and Supp. Fig. S6). Interestingly, this decrease in gradient sensitivity does not require receptor saturation, as is commonly assumed (44). Calculations that account for true receptor geometries of the sensor give results similar to Eq. [1] (6). A critical prediction of this theory is that precision of gradient sensing (expressed as SNR) always increases with the organoid size $A$. We next contrasted this prediction with experimental data.

To examine whether the precision of gradient sensing increases with the organoid size, we examined the bias of response of differently sized organoids naturally formed in our assays. (Fig. 3). To enable the comparison, we computed the fraction of organoids with $L_U > L_D$, where $L_U(L_D)$ is the sum of branch lengths (projected in the gradient direction) pointing up (down) the gradient (measure B in Fig. 2). The corresponding theoretical prediction can be inferred from the analysis of a one-dimensional array of $N$ coupled cells subjected to a ligand gradient. In particular the experimentally determined difference between 'up' and 'down' pointing branch numbers can be compared with the theoretically predicted probability that the measured number of ligand molecules in the $N$'th cell is larger than in the first cell in the array, $\nu_N > \nu_1$. We take $\nu_n$ as Gaussian-distributed with mean $\bar{c}_n a^3$ and variance $\bar{c}_n a^3 + \eta^2$, where the first term accounts for the Poisson nature of the molecular counts, and $\eta^2$ represents the additional noise downstream of sensing, which can dominate the sensory noise, but is assumed to be unbiased (multiplicative noise was also considered, with similar effects, see *SI* and Supp. Fig. S7). We set the value of $\eta^2$ by equating the experimental and theoretical bias probabilities averaged over all organoid sizes and background concentrations observed in the experiments. Figure 3A demonstrates that bias increases roughly linearly with the gradient strength in both the experiments and the BP model. However, Fig. 3B shows that the experimental bias saturates with organoid size, while the BP theory would predict an increase without bounds. Further, Fig. 3C shows that the experimental bias is generally weaker than that predicted by the BP theory. These disagreements with experimental results suggest that a new theory of multicellular gradient detection is required.

---

[1] This is similar to the observation that a small difference of two large numbers always has a larger relative error than either of the two numbers, and so one is frequently cautioned against making such subtractions in scientific computing (43).

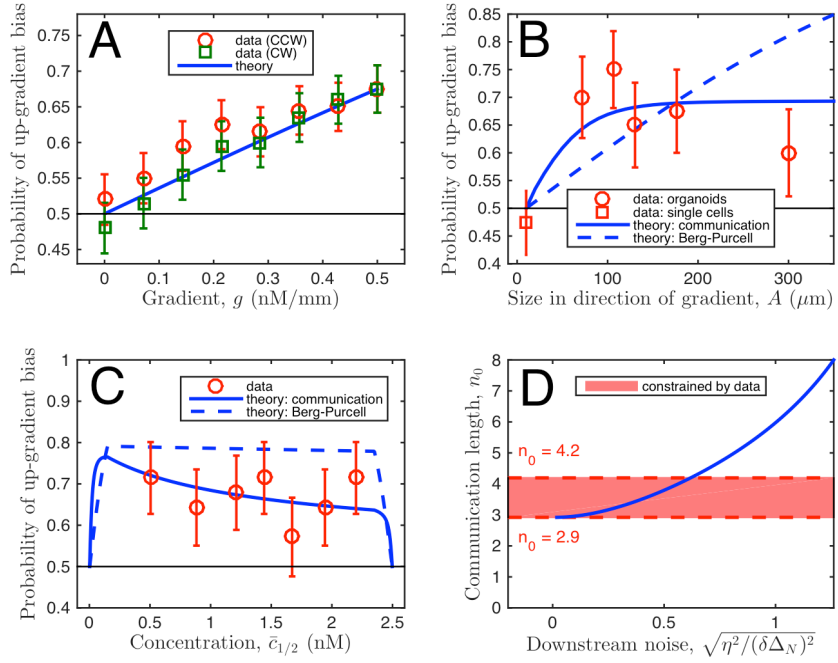

**Figure 3. Organoid branching bias for different parameters: data support theory of communication-constrained gradient sensing.** (A-B) Experimentally measured bias (denoted as 'data') is plotted vs.: (A) EGF gradient strength, (B) organoid size, and (C) background EGF concentration. For the data, bias is measured as the fraction of organoids with $L_U > L_D$, where $L_U (L_D)$ is the sum of branch lengths, projected in the gradient direction, pointing up (down) the gradient (measure B in Fig. 2). Error bars are standard errors. In (A), gradient response is estimated for different branching response directions, with the response axis rotated by the angle $\theta$, either clockwise (CW) or counterclockwise (CCW) with respect to the gradient directions, reducing the effective gradient by $\cos\theta$. In (B), the organoid body size is measured excluding branches, and the bias for single cell migration is measured as for organoid branches, but with cell displacement replacing branch length. In (C), stability of bias at large concentration shows absence of significant saturation effects. *Theory:* The bias is calculated as the probability that $\Delta_N > \Delta_1$, where $\Delta_N (\Delta_1)$ is the Gaussian-distributed concentration difference in the cell furthest up (down) the gradient, with mean and variance given by the model as described in the text. The classic Berg-Purcell (BP) theory predictions, which do not account for communication, are equivalent to setting the communication length $n_0$ to infinity in the theory that accounts for the communication. The downstream noise $\eta^2$ is set by the data. In A, this leaves no free parameters, and in B and C, after assuming a cell size of $a = 10$ $\mu$m and taking the limit of large gain $\beta/\mu$, this leaves only the communication length $n_0$. The curve for $n_0 = 3.5$ is shown in comparison to the BP curve to illustrate the effect of communication. In (B) and (C), the theoretical curves are obtained by averaging uniformly over the ranges of organoid size or midpoint concentration seen in the experiments (10-600 $\mu$m, 0.22-2.47 nM), except where prohibited by geometry, as in Supp. Fig. S6. Since background concentration and size of the organoids are not independent, this results in the nonmonotonic dependence in (C). Specifically, the nonmonotonicity comes from the noisy comparison of large, similar concentrations, as explained in the text, and also because organoids with large midpoint concentration are at the edge of the device and cannot be large. (D) Communication length $n_0$ is constrained by the data. Minimum $n_0 \geq 2.9$ is determined by the fact that downstream processes can only add noise, not remove it ($\eta^2 \geq 0$). Maximum $n_0 \leq 4.2$ is determined by the condition that data and theory in B agree sufficiently that $\chi^2$/(degree of freedom) < 1.

To develop the new theory, we note that, by assuming that information, collected by different parts of a spatially extended detector, can be integrated in an essentially error free fashion, the BP approach neglects a major complication: the communication noise. Indeed, to contrast spatially distributed inputs, e.g., the local EGF concentration, the information collected in different parts of a coupled multi-cellular ensemble must be communicated over large distances by means of noisy, molecular diffusion and transport processes. The unavoidable communication errors set new, unknown limits on the highest accuracy of sensing. From this perspective, the BP analysis accounts for the *extrinsic* noise of the ligand concentration, but not for the *intrinsic* noise (3, 45) of multi-cellular communication. To study the communication noise effects, we again approximated an organoid by a one-dimensional chain of $N$ cells, each of size $a$, for a total length of $A=Na$ parallel to the gradient direction. The observed independence of the response bias of the background EGF concentration (Fig. 3C) supports an adaptive model of sensing. We chose a minimal adaptive model allowing for chemical diffusive communication, based on the principle of local excitation and global inhibition (LEGI) (8, 46, 47). In the $n$th cell, both a local and a global molecular messenger species are assumed to be produced in proportion to the local external EGF concentration $c_n$ at a rate $\beta$, and are degraded at a rate $\mu$. Whereas the local messenger species is confined to each cell, the global messenger species is exchanged between neighboring cells at a rate $\gamma$, which provides an intrinsically noisy communication. The local messenger then excites a downstream species, while the global messenger inhibits it. In the limit of shallow gradients, the excitation level reports the difference $\Delta_n$ between local and global species concentrations (see *SI*). The difference $\Delta_{N,1}$ in the edge cells provide the sensory readout: positive/negative $\Delta$ shows that the local concentration at the edge is above/below the average, and hence the cell is up/down the gradient. Note that an individual cell within this multi-cellular version of the LEGI model cannot detect a gradient, as the readout will always be zero within statistical fluctuations.

In our analysis, we again note the absence of temporal integration of EGF gradients (Supp. Fig. S5; see also the extension of our analysis to the temporal integration case in Ref. (41)). Further, since there is no evidence for receptor saturation at high concentration (Fig. 3C), we confine ourselves to the linear response regime for theoretical studies. These assumptions allow us to calculate the limit of the sensory precision of the gradient detection, as a function of organoid size $N$ and the background concentration (see *SI* and Supp. Fig. S6). We find that precision initially grows with $N$, then saturates at a maximal value (Supp Fig. S6C). This is in contrast to the BP estimate, Eq. [1], which predicts that precision grows indefinitely with $N$. In our model we expect precision to be the highest in the limits of a large organoid ($N \gg 1$), fast cell-to-cell communication ($\gamma/\mu \gg 1$), and large local and global messenger species concentrations ($\beta/\mu \gg 1$). In these limits the saturating value of

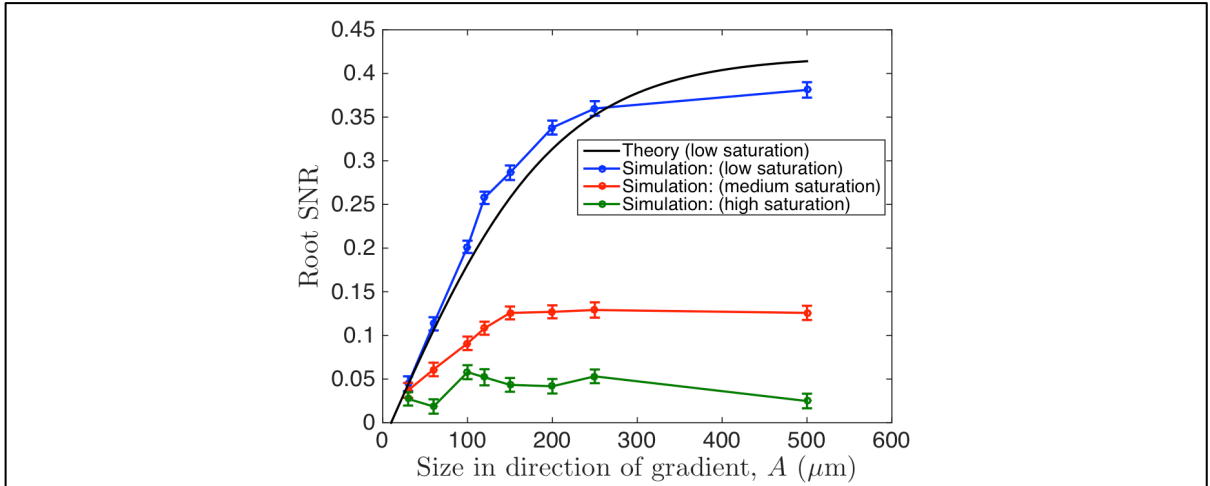

**Figure 4. Saturation of signaling responses reduces the maximum SNR in the simulation of multicellular gradient sensing.** To investigate effects of signaling saturation, we simulated a multicellular stochastic LEGI model using a spatially resolved implementation of the stochastic simulation algorithm (see *SI*) and measured the simulated organoid bias at different saturation levels. The figure shows the square root of the SNR (mean response squared over its variance), where for the response we take the deviation of the species activated by the local and suppressed by the global messenger from its mean value. This is the closest equivalent to $\Delta_N$, and to the bias measure in Fig. 3 and Supp. Fig. S6. For the theory and the simulations, we set $g = 5$ molecules per cell volume per cell diameter, $n_0 = \sqrt{\gamma/\mu} = 10$ cells, $\bar{c}_N = 1000$ molecules per cell volume. We contrasted the theory and the simulations by matching the low saturation curves for small organoid sizes. The effects of saturation in activation/suppression of the response by the messengers result in earlier saturation of the SNR curves, as expected.

the sensory error takes the simplified form (Eq. [46] in *SI*):

$$\frac{1}{\text{SNR}} = \left(\frac{\delta \Delta_N}{\bar{\Delta}_N}\right)^2 = \frac{\bar{c}_N}{a^3(n_0 ag)^2} = \frac{\bar{c}_{1/2} + gaN/2}{a^3(n_0 ag)^2}, \quad [2]$$

where $n_0 = \sqrt{\gamma/\mu}$. Comparing Eq. [2] to the BP estimate, Eq. [1], we see that even when communication noise is accounted for, the organoid can achieve the noise-free bound, but with an *effective size* of $A = n_0 a$ instead of the actual size $A = Na$. Thus $n_0$, which grows with the communication rate $\gamma$, sets the length scale of the effective sensory unit within the organoid: it is the number of neighbors with which a cell can reliably communicate before the information becomes degraded by the noise. Beyond $N \sim n_0$, a larger organoid is predicted to achieve no further benefit to its sensory precision. Additionally, because of this finite communication length scale, the sensory precision is predicted to depend on the concentration at the edge cell(s), rather than in the middle of the organoid. Thus the interplay between the concentration and the organoid size is also very different compared to the predictions of the standard BP theory.

We first tested the new theory that accounts for communication by simulating the multi-cellular, LEGI-based sensing with a spatially extended Gillespie algorithm (see *SI* for details). This analysis allowed us to explore the non-linear (Michaelis-Menten type) biochemical reaction regime. We verified that our theoretical predictions were fully consistent with this stochastic model in the linear regime, and were still qualitatively valid when the dependence of the local and the global signaling reactions on the input was allowed to gradually saturate (Fig. 4). In particular, under all assumptions, the advantage of increasing detector rapidly reached a maximum value. This maximum SNR value, however, gradually decreased with increasing saturation, suggesting predominant effects of decreasing sensitivity of saturating chemical reactions to the differences in the input values.

We then compared the predictions of our new theory of multicellular gradient sensing to the experimental measurements. To do that, we calculated the probability that the gradient indeed biases the branching response, i.e., that $\Delta_N > \Delta_1$, where $\Delta_n$ was assumed in the theory to be a Gaussian-distributed variable with the mean $\bar{\Delta}_n$ and variance $(\delta\Delta_n)^2 + \eta^2$ (the case of the multiplicative noise is treated in the SI and Supp. Fig. S7). The first term in the variance is calculated in the *SI*, and the second reflects the added noise downstream of gradient sensing, set by the average organoid bias, identical to the one found in the BP theory above. Figure 3A-C demonstrates the excellent agreement between experiment and theory that accounts for the communication noise, suggesting that the new theory is a much better explanation of the data than the BP analysis.

The experimental data in Fig. 3B place constraints on the possible range of values of the size of the effective multi-cellular sensing unit, $n_0$. The requirement that $\eta^2 \geq 0$ (downstream processes only increase noise, they do not decrease it) places the lower bound $n_0 \geq 2.9$ (Fig. 3D). Roughly speaking, the edge cell must communicate with at least three neighbors if the inability of the observed bias to reach 1 was due *entirely* to sensory noise, with no additional noise downstream. Further, the requirement that the model agree with the data within error bars in Fig. 3B also places the upper bound $n_0 \leq 4.2$ (similar limits come from the multiplicative model, Supp. Fig. S7). That is, a functional sensing unit of four cells or less is required to explain why all organoids, which range in width from approximately 8 to 30 cells, display roughly the same bias, independent of their size. Thus Fig. 3 demonstrates that cells receive reliable information from only a few nearby cells, and this number is tightly bounded. The tightness of the bound implies that the noise downstream of the sensing process is relatively small. Crucially, in our theory, a cell not communicating with the neighbors cannot detect a gradient, and a nonzero value of $n_0$ is *qualitatively* different from $n_0 = 0$. We thus tested if gradient sensing would be altered if cell-cell communication was prevented in the organoids.

A central prediction of the theoretical analysis is that preventing cell-cell communication can lead to a complete loss of sensing of shallow gradients. One simple way cell-cell communication can occur in epithelial layers is by means of gap junctions. We therefore explored the effect of disrupting the gap junction communication using four distinct inhibitors: 50 nM Endothelin-1, 50 μM flufenamic acid, 0.5 mM octanol and μM carbenoxolone (48). Although the mode of inhibition was different for these distinct compounds, application of each one of them resulted in a complete loss of directional bias in response, while the branching itself was present, and was similar to that without gap junction perturbation in spatially uniform EGF concentrations (Fig. 5A, Supp. Fig. S8). Crucially, this result also confirms that communication over the effective sensory unit is due to intracellular chemical diffusion, rather than through the extracellular medium or due to a mechanical coupling. The likely candidates for gap junction mediated cell-cell coupling are calcium or inositol trisphosphate (IP3), both of which are second messengers that can control intracellular Ca release. EGF is known to stimulate Ca signaling (49) at least in part through stimulation of IP3 synthesis, thus providing a source of these intracellular messengers.

To examine Ca signaling more directly, we used organoids obtained from transgenic mice, expressing genetically encoded Ca reporter GCaMP4, under the control of the CAG promoter (50). We confirmed that addition of 2.5 nM EGF to the medium indeed triggered a pulse of calcium signaling in a typical organoid (Fig. 5C). Furthermore, the Ca activity throughout the branching processes was coordinated, releasing calcium nearly simultaneously in cells at the tips of growing branches, suggesting cell-cell communication leading to Ca releases (see *SI*, Supp. Fig. S9, and Supp. Movie 1). To deplete intracellular Ca stores and thus potentially disrupt the effect of chemical cell-cell coupling, we treated the organoids with sarco/endoplasmic reticulum Ca2+-ATPase (SERCA) inhibitor thapsigargin. This treatment indeed was sufficient to disrupt EGF gradient sensing in the treated organoids (Fig. 5B and Supp. Fig. S2D). Surprisingly, SERCA inhibition also enhanced the branching elongation: the average length of a branch increased from $74 \pm 1 \mu m$ for WT organoids to $201 \pm 3 \mu m$ with SERCA blocking; the organoids appear to be almost entirely composed of branches after 3 days under these conditions. This result suggested that gap junction-mediated exchange of a molecular regulator that can trigger intracellular calcium release may have a negative effect on the local branching response, consistent with the assumed negative role of the diffusive messenger postulated in the LEGI model. We finally note that small molecules exchanged though gap junctions (e.g., IP3 or calcium ions) would be a natural choice for the cell-cell coupling intercellular messenger, since their smaller size and larger diffusion coefficient (compared to peptides) allow for a larger $\gamma$, which, in turn, increases the size of the effective sensory unit $n_0$ and improves the sensing accuracy.

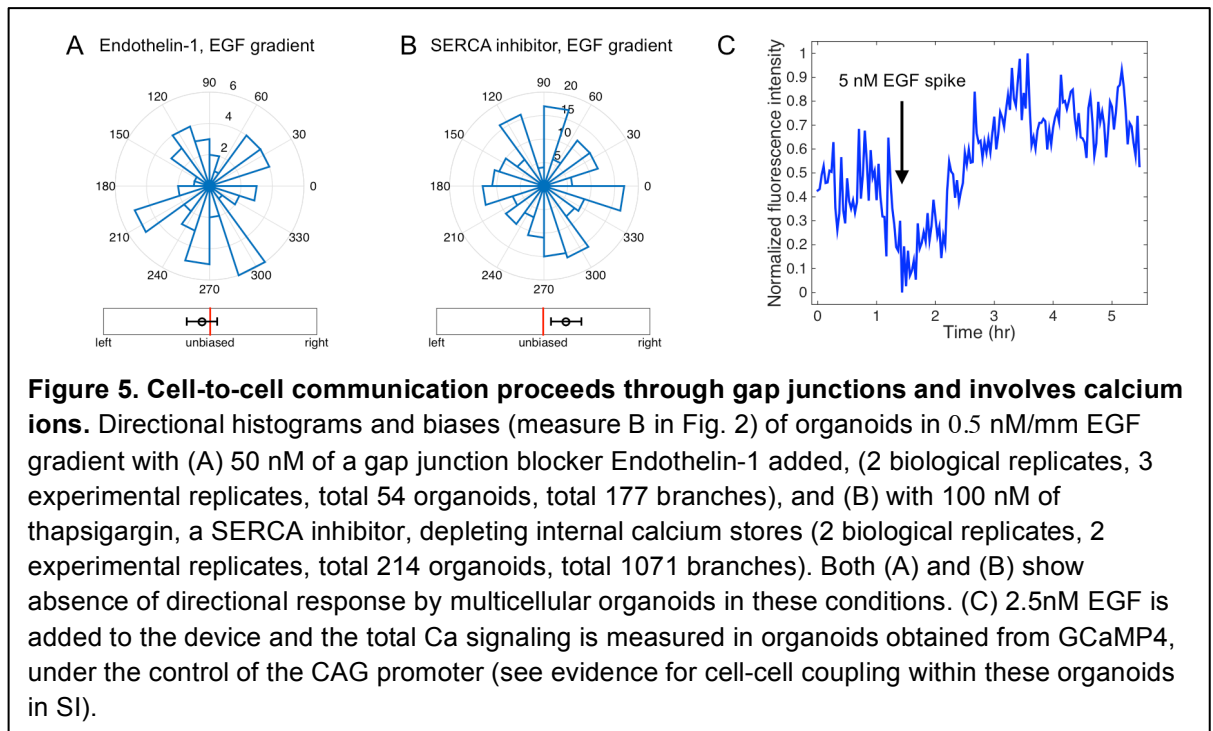

**Figure 5. Cell-to-cell communication proceeds through gap junctions and involves calcium ions.** Directional histograms and biases (measure B in Fig. 2) of organoids in $0.5$ nM/mm EGF gradient with (A) 50 nM of a gap junction blocker Endothelin-1 added, (2 biological replicates, 3 experimental replicates, total 54 organoids, total 177 branches), and (B) with 100 nM of thapsigargin, a SERCA inhibitor, depleting internal calcium stores (2 biological replicates, 2 experimental replicates, total 214 organoids, total 1071 branches). Both (A) and (B) show absence of directional response by multicellular organoids in these conditions. (C) 2.5nM EGF is added to the device and the total Ca signaling is measured in organoids obtained from GCaMP4, under the control of the CAG promoter (see evidence for cell-cell coupling within these organoids in SI).

## Discussion

Morphogenesis and growth of complex tissues is orchestrated by diverse chemical and mechanical cues. These cues not only specify patterning of developing tissues but also direct tissue growth and expansion. However, we still lack details of how these collective, multi-cellular processes are controlled by spatial gradients of extracellular ligand molecules. Here we used mathematical modeling, computational simulations, and experimentation in a novel gradient generating device to study the directional guidance of branch formation and extension in a model of mammary tissue morphogenesis. Our data revealed that multicellular constructs undergo directionally biased migration in shallow gradients of EGF that are undetectable to single cells. Further, our analysis suggests that cell-cell communication through gap junctions underlies the increased gradient sensitivity, allowing the cell ensembles to expand the range of EGF concentrations they can sense within the gradient, and thus enhance the overall guidance signal. Increasing evidence suggests that collective sensing of environmental signals, particularly if accompanied by secretion of a common signal that enables averaging of variable and noisy signaling in individual cells, can help improve reliability of signaling, cell fate choices, and behavioral actions. Examples are abundant in coordinated pathogen actions or immune responses [8, 37-43]. Similarly, individual sensing and collective decision-making in morphogenesis and animal group behaviors have been shown to amplify weak signals observed by individual agents and to develop coherent, long-range patterns (24, 25, 51, 52). In contrast to 'all-to-all' signaling or response communication cases, here we focused on the case of sequential communication of a signal between the sensing units, in a relay fashion, which can enhance the sensing precision by enhancing the effective input itself. Critically, this communication mechanism, mediated by diffusive coupling through gap junctions, can be seen as an information-theoretic relay channel (53, 54), see Fig. 6. The theoretical analysis we present here is thus one of the first departures from the simple point-to-point information-processing paradigm in systems biology. In fact, our calculations of reliability of multicellular signaling, presented in this paper and in (41), are equivalent to calculating channel capacities of various Gaussian relay channels.

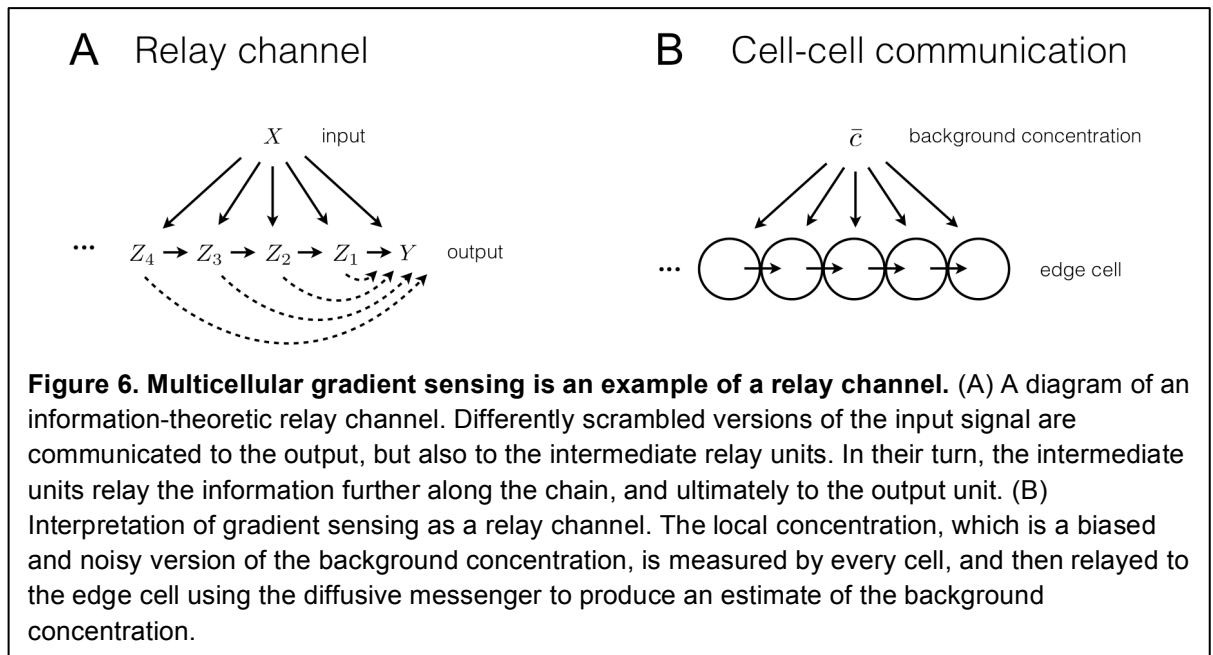

**Figure 6. Multicellular gradient sensing is an example of a relay channel.** (A) A diagram of an information-theoretic relay channel. Differently scrambled versions of the input signal are communicated to the output, but also to the intermediate relay units. In their turn, the intermediate units relay the information further along the chain, and ultimately to the output unit. (B) Interpretation of gradient sensing as a relay channel. The local concentration, which is a biased and noisy version of the background concentration, is measured by every cell, and then relayed to the edge cell using the diffusive messenger to produce an estimate of the background concentration.

The key consequence of the relay communication mechanism is that it is subject to a gradual buildup of communication noise, mitigating the gain from the signal increase, and providing a fundamental limit on effectiveness of such collective sensing responses. This result runs counter to the prevailing intuition that sensing accuracy should increase without bound with the system size (40), for multicellular systems in development (27) and also for other multi-agent sensory systems. These intuitive expectations are flawed precisely because they fail to take into account the importance of communication uncertainty, which provides fundamental limits on the gains resulting from multicellular sensing. Our integrated analysis reveals that this multicellular sensing strategy in growing mammary branches is indeed limited by the noisy cell-cell communication. Importantly, we were able to combine theory and experiments to estimate these limits for EGF gradient response of mammary branching and found them to be much tighter than those that assume that all of the spatially distributed information is immediately actionable: growth of the branch beyond the size of the maximum effective multicellular sensing unit does not improve the sensing accuracy. We estimate that the sensing unit is approximately 3-4 cell lengths, a size that is consistent with the number of cell layers in small end buds of a growing mammary duct (55) (see also Supp. Movie 1). Some large end buds in vivo contain significantly more cell layers and our analysis suggests that these additional cells may be primarily involved in other functions, such as proliferation and differentiation, and not gradient sensing. The narrow bounds on the number of interacting cells also suggest that the "actuation" noise downstream of sensing is minimal, paralleling related findings in the nervous system (56). Interestingly, the theoretical analysis predicted that the sensory unit size is specified by a simple formula describing the typical distance traveled by a diffusing messenger molecule before it degrades or is inactivated, consistent with simpler estimates of the molecular communication reach (57). Our analysis also provided a new way to interpret the dependence between the background ligand concentration and gradient sensing – saturation of receptors is not needed to explain the often-observed decrease in the sensory precision at high concentration (12, 13, 44). Rather, the loss of precision is ascribed to increasing noise to signal ratio, stemming from the need to compare large, noisy concentrations. Similar limits might exist in any biological systems with spatially distributed sensing of spatially graded signals, including single cells or multi-nuclear syncytia.

Our results suggest that the intercellular communication underlying multicellular sensing in growing mammary tissue is mediated by calcium signaling events, as depletion of internal stores by a SERCA inhibitor both enhanced the branch formation and inhibited gradient detection. Thus release of calcium from internal stores is consistent with a negative or limiting effect on the local branch formation or extension. The release can be controlled by either IP3 or calcium itself, both of which can diffuse through gap junctions. Therefore, the inhibitory diffusive signal postulated by the LEGI models of gradient sensing may rely on the ultimate release of calcium from internal stories, as also suggested by our imaging of calcium with the genetically encoded probe. This role of calcium is consistent with its enhancement of retraction of the leading front in migrating cells (58). Consistent with the LEGI model, gradient sensing was persistent in time and exhibited very low sensitivity to the local background EGF concentration. The use of the LEGI model in our analysis, both in mathematical modeling and in spatially distributed Gillespie simulations, also showed results quantitatively consistent with the experiments, suggesting that this model was appropriate for describing the diffusively coupled collective EGF sensing.

Overall, we conclude that collective gradient sensing suggested for many natural developmental processes (59), as well as for pathological invasive tissue expansion (37), is an effective strategy, which, though subject to important limitations, can help explain the observed differences in the single cell and multicellular chemotactic responses. Importantly, the experimentally validated theory proposed in our analysis provides a way to assess the potential role of inter-cellular communication in other settings, including invasive tumor growth, pointing to the specific parameters that can be altered to disrupt this process or make it less efficient.

## Materials and Methods

*Experimental device.* Custom PDMS devices were developed using stereolithography, yielding culture area approximately 5mm wide, 10mm long and 1mm tall (see Fig. 1B). The sides of the device are open wells that allow the use of standard pipettes to change media and six replicates of the entire device is contained within a standard six-well plate.  Before use, the center cell culture area is filled.  This action is assisted by the hexagonal pillars, which are used to trap the liquid 3D ECM and organoid mixture within the cell culture area before the 3D ECM can harden (60). Once the 3D ECM matrix of choice has hardened, the open wells can be filled as previously mentioned.  Both *in silico* and *in vivo* (see *SI* and Supp. Fig. S10) tests demonstrate a stable linear EGF gradient across the cell culture area for approximately three days, after which the media can be replenished as needed. Various compounds were added to collagen gel at final concentrations, as indicated, along with the organoids.

*Stereolithography & PDMS Casting.* Using the 3D rendering software SolidWorks (Dassault Systems), we drew the final mold for the PDMS devices.  The design was electronically transmitted to FineLine Prototyping (Raleigh, NC) where it was rendered using high-resolution ProtoTherm 12120 as the material with a natural finish.  Proprietary settings were used to accurately render the pillars.  In two to three days the mold was shipped and after its arrival we mixed PDMS monomer to curing agent in a 10:1 ratio (Momentive RTV615).  After mixing, the liquid PDMS was poured into the mold and a homemade press was used to keep the top surface flat.  This press from bottom to top consisted of a steel plate, paper towel, piece of a clear transparency film, the mold with PDMS, another piece of a transparency, paper towel, piece of rubber, and an another steel plate.  The entire assembly was placed in the oven at 80oC and baked overnight. The devices were then washed, cut, and placed on top of 22x22mm coverslips (72204-01, Electron Microscopy Sciences). Six devices where then placed inside an autoclave bag and sterilized. When needed, the bag was opened in a sterile environment and the devices were filled and placed inside a 6-well plate.

*Device preparation for time-lapse imaging.* In order to allow for real-time imaging, the devices were fabricated as described above, and were then cut from the PDMS using a 16mm sharp leather punch to create a circular device. The device was sterilized with ethanol and then plasma treated before being bonded directly to the bottom of a glass-bottomed 6-well plate with a 20mm hole (LiveAssay).

*Collagen preparation.* Rat-tail Collagen (354236, BD Biosciences) was pH balanced using 1M NaOH (S2770, Sigma) mixed to a final concentration of 3mg/mL with 10x DMEM (D2429, Sigma). This mixture sat in an ice block in aliquots of no more than 1.5mL until fibers formed, typically approximately 75 min (described in detail in Ref. (33)). Cells were then mixed in at 2.5 organoids/mL and a 100µL pipet tip was used to draw 75µL of the suspension. The pipette was inserted into the pre-punched hole and the suspension was gently injected into the device. The device was placed on a heat block for no more than 10 min before the side wells were filled with solution. The lid was replaced on the 6-well plate and the whole assembly was plated inside the incubator at 37oC with 5% $CO_2$.

*Confocal Microscopy.* Confocal imaging was performed as previously reported (32, 61). Briefly, imaging was done with a Solamere Technology Group spinning disk confocal microscope, using a 40× C-Apochromat objective lens (Zeiss Microimaging). Both fixed and time-lapse images were acquired using a customized combination of µManager (https://www.micro-manager.org) and Piper (Stanford Photonics). Thereafter image stacking and adjustments were done with Imaris (Bitplane) in order to maximize clarity, but these adjustments were always done on the entire image.

*DIC Microscopy.* Phase contrast images were taken with an Axio Observer DIC inverted microscope (Carl Zeiss, Inc) using AxioVision Software (Carl Zeiss, Inc). All image processing was either done with Adobe Photoshop CS 6.0 (Adobe) or Fiji (GPL v.2) for clarity, but always done on the entire image.

*Image Quantification.* A custom Fiji program was written to measure the angle and length of the resulting branches. Additionally this program allows the user to draw a freehand outline around the body and/or the branch and body of the whole organoid. From these outlines area, a fit ellipse, and a Feret diameter were computed along with related statistics. After all measurements were made, a custom MatLab (MatWorks, Inc) program was written to create the graphs.

*Primary mammary organoid isolation.* Cultures are prepared as previously described (33). Mammary glands are minced and tissue is shaken for 30 min at 37°C in a 50 ml collagenase/trypsin solution in DMEM/F12 (GIBCO-BRL), 0.1 g trypsin (GIBCO-BRL), 0.1 g collagenase (Sigma C5138), 5 ml fetal calf serum, 250 µl of 1 µg/ml insulin, and 50 µl of 50 µg/ml gentamicin (all UCSF Cell Culture Facility). The collagenase solution is centrifuged at 1500 rpm for 10 min, dispersed through 10 ml DMEM/F12, centrifuged at 1500 rpm for 10 min, and then resuspended in 4 ml DMEM/F12 + 40 µl DNase (2U/µl) (Sigma). The DNase solution is shaken by hand for 2–5 min, then centrifuged at 1500 rpm for 10 min. Organoids are separated from single cells through four differential centrifugations (pulse to 1500 rpm in 10 ml DMEM/F12). The final pellet is resuspended in the desired amount of Growth Factor Reduced collagen.

*Multicellular gradient sensing model.* Theoretical results are derived using a stochastic dynamical model of multicellular sensing and communication. The model includes Langevin-type noise terms corresponding to ligand number fluctuations, stochastic production and degradation of internal messenger molecules, and exchange of messenger molecules between neighboring cells in a one-dimensional chain. The model is linearized around the steady state. The mean and instantaneous variance of the readout variable $\Delta N$ are obtained by Fourier transforming and integrating the power spectra over all frequencies. This leads to an expression in terms of the matrix of exchange reactions, whose inverse (the "communication kernel") we solve for analytically and approximate in the appropriate limits to obtain Eq. [2]. See *SI* for more information.

*Statistical Analysis.* Angular histograms (e.g., Fig. 1C-F) plot the distribution of branching directions over all organoids. For each organoid, the branching direction is defined as the angle of the vector sum of its branches. A branch vector extends from the organoid body (defined by the fitted ellipse) to the tip of the branch. For single cell movement (Fig. 1E), the definitions are the same, except that branch vector is replaced by the displacement vector, from where the cell broke away from the organoid, to where the cell is observed in the image. The breakaway point is taken to be the nearest branch tip. Data contained in the angular histograms are reduced to a single bias measure in one of six ways, as described in Fig. 2**.** Measure B is also shown in Fig. 1, 3 and 4. See *SI* for comparison of the bias measures.

## Acknowledgements

We thank Peng Huang for useful discussions. This work was supported in part by James S. McDonnel Foundation grant No. 220020321 (AM and IN), by NSF grants No. 1410978 (IN), 1410593 (AE), and 1410545 (AL), and NIH grant GM072024 (AL).

## References

1. Swaney KF, Huang CH, & Devreotes PN (2010) Eukaryotic chemotaxis: a network of signaling pathways controls motility, directional sensing, and polarity. *Annual review of biophysics* 39:265-289.
2. Levchenko A & Nemenman I (2014) Cellular noise and information transmission. *Current opinion in biotechnology* 28:156-164.
3. Swain PS, Elowitz MB, & Siggia ED (2002) Intrinsic and extrinsic contributions to stochasticity in gene expression. *Proc Natl Acad Sci U S A* 99(20):12795-12800.
4. Munsky B, Neuert G, & van Oudenaarden A (2012) Using gene expression noise to understand gene regulation. *Science* 336(6078):183-187.
5. Cheong R, Rhee A, Wang CJ, Nemenman I, & Levchenko A (2011) Information transduction capacity of noisy biochemical signaling networks. *Science* 334(6054):354-358.
6. Endres RG & Wingreen NS (2008) Accuracy of direct gradient sensing by single cells. *Proc Natl Acad Sci U S A* 105(41):15749-15754.
7. Andrews BW & Iglesias PA (2007) An information-theoretic characterization of the optimal gradient sensing response of cells. *PLoS Comput Biol* 3(8):e153.
8. Jilkine A & Edelstein-Keshet L (2011) A comparison of mathematical models for polarization of single eukaryotic cells in response to guided cues. *PLoS Comput Biol* 7(4):e1001121.
9. Hu B, Chen W, Rappel WJ, & Levine H (2010) Physical limits on cellular sensing of spatial gradients. *Phys Rev Lett* 105(4):048104.
10. Hu B, Chen W, Levine H, & Rappel WJ (2011) Quantifying information transmission in eukaryotic gradient sensing and chemotactic response. *J Stat Phys* 142(6):1167-1186.
11. Fuller D*, et al.* (2010) External and internal constraints on eukaryotic chemotaxis. *Proc Natl Acad Sci U S A* 107(21):9656-9659.
12. Goodhill GJ & Urbach JS (1999) Theoretical analysis of gradient detection by growth cones. *Journal of neurobiology* 41(2):230-241.
13. Rosoff WJ*, et al.* (2004) A new chemotaxis assay shows the extreme sensitivity of axons to

molecular gradients. *Nature neuroscience* 7(6):678-682.
14. Chung CY, Funamoto S, & Firtel RA (2001) Signaling pathways controlling cell polarity and chemotaxis. *Trends in biochemical sciences* 26(9):557-566.
15. Takahashi Y, Sipp D, & Enomoto H (2013) Tissue interactions in neural crest cell development and disease. *Science* 341(6148):860-863.
16. Dona E*, et al.* (2013) Directional tissue migration through a self-generated chemokine gradient. *Nature* 503(7475):285-289.
17. Pocha SM & Montell DJ (2014) Cellular and molecular mechanisms of single and collective cell migrations in Drosophila: themes and variations. *Annual review of genetics* 48:295-318.
18. Lu P & Werb Z (2008) Patterning mechanisms of branched organs. *Science* 322(5907):1506-1509.
19. Friedl P & Gilmour D (2009) Collective cell migration in morphogenesis, regeneration and cancer. *Nat Rev Mol Cell Biol* 10(7):445-457.
20. Gregor T, Fujimoto K, Masaki N, & Sawai S (2010) The onset of collective behavior in social amoebae. *Science* 328(5981):1021-1025.
21. Malet-Engra G*, et al.* (2015) Collective cell motility promotes chemotactic prowess and resistance to chemorepulsion. *Current biology : CB* 25(2):242-250.
22. Lin B, Yin T, Wu YI, Inoue T, & Levchenko A (2015) Interplay between chemotaxis and contact inhibition of locomotion determines exploratory cell migration. *Nat Commun* 6:6619.
23. Camley B, Zimmermann J, Levine H, & Rappel WJ (2015) Emergent collective chemotaxis without single-cell gradient sensing. *arXiv* 1506.06698
24. Taillefumier T & Wingreen NS (2015) Optimal census by quorum sensing. *PLoS Comput Biol* 11(5):e1004238.
25. Voisinne G*, et al.* (2015) T Cells Integrate Local and Global Cues to Discriminate between Structurally Similar Antigens. *Cell reports*.
26. Sokolowski T & Tkacik G (2015) Optimizing information flow in small genetic networks. IV. Spatial coupling. *arXiv* 1501.04015.
27. Lander AD (2011) Pattern, growth, and control. *Cell* 144(6):955-969.
28. Inman JL, Robertson C, Mott JD, & Bissell MJ (2015) Mammary gland development: cell fate specification, stem cells and the microenvironment. *Development* 142(6):1028-1042.
29. Huebner RJ & Ewald AJ (2014) Cellular foundations of mammary tubulogenesis. *Seminars in cell & developmental biology* 31:124-131.
30. Sternlicht MD, Kouros-Mehr H, Lu P, & Werb Z (2006) Hormonal and local control of mammary branching morphogenesis. *Differentiation; research in biological diversity* 74(7):365-381.
31. Fata JE*, et al.* (2007) The MAPK(ERK-1,2) pathway integrates distinct and antagonistic signals from TGFalpha and FGF7 in morphogenesis of mouse mammary epithelium. *Developmental biology* 306(1):193-207.
32. Ewald AJ, Brenot A, Duong M, Chan BS, & Werb Z (2008) Collective epithelial migration and cell rearrangements drive mammary branching morphogenesis. *Developmental cell* 14(4):570-581.
33. Nguyen-Ngoc KV*, et al.* (2015) 3D culture assays of murine mammary branching morphogenesis and epithelial invasion. *Methods Mol Biol* 1189:135-162.
34. Schmeichel KL & Bissell MJ (2003) Modeling tissue-specific signaling and organ function in three dimensions. *J Cell Sci* 116(Pt 12):2377-2388.
35. Coleman S, Silberstein GB, & Daniel CW (1988) Ductal morphogenesis in the mouse mammary gland: evidence supporting a role for epidermal growth factor. *Developmental biology* 127(2):304-315.
36. Sebastian J*, et al.* (1998) Activation and function of the epidermal growth factor receptor and erbB-2 during mammary gland morphogenesis. *Cell growth & differentiation : the molecular biology journal of the American Association for Cancer Research* 9(9):777-785.
37. Roussos ET, Condeelis JS, & Patsialou A (2011) Chemotaxis in cancer. *Nature reviews. Cancer* 11(8):573-587.
38. Radice GL*, et al.* (1997) Precocious mammary gland development in P-cadherin-deficient mice. *The Journal of cell biology* 139(4):1025-1032.
39. Nguyen-Ngoc KV*, et al.* (2012) ECM microenvironment regulates collective migration and local dissemination in normal and malignant mammary epithelium. *Proc Natl Acad Sci U S A* 109(39):E2595-2604.
40. Berg HC & Purcell EM (1977) Physics of chemoreception. *Biophys J* 20(2):193-219.
41. Mugler A, Levchenko A, & Nemenman I (2015) Limits to the precision of gradient sensing

with spatial communication and temporal integration. *arXiv*:1505.04346
42. Berg HC (2003) *E. Coli in motion* (Springer).
43. Press WH (2007) *Numerical recipes : the art of scientific computing* (Cambridge University Press, Cambridge, UK ; New York) 3rd Ed pp xxi, 1235 p.
44. Mortimer D, Dayan P, Burrage K, & Goodhill GJ (2011) Bayes-optimal chemotaxis. *Neural Comput* 23(2):336-373.
45. Elowitz MB, Levine AJ, Siggia ED, & Swain PS (2002) Stochastic gene expression in a single cell. *Science* 297(5584):1183-1186.
46. Iglesias PA & Levchenko A (2002) Modeling the cell's guidance system. *Sci STKE* 2002(148):RE12.
47. Levchenko A & Iglesias PA (2002) Models of eukaryotic gradient sensing: application to chemotaxis of amoebae and neutrophils. *Biophys J* 82(1 Pt 1):50-63.
48. Salameh A & Dhein S (2005) Pharmacology of gap junctions. New pharmacological targets for treatment of arrhythmia, seizure and cancer? *Biochimica et biophysica acta* 1719(1-2):36-58.
49. Dittmar T, *et al.* (2002) Induction of cancer cell migration by epidermal growth factor is initiated by specific phosphorylation of tyrosine 1248 of c-erbB-2 receptor via EGFR. *FASEB journal : official publication of the Federation of American Societies for Experimental Biology* 16(13):1823-1825.
50. Zariwala HA, *et al.* (2012) A Cre-dependent GCaMP3 reporter mouse for neuronal imaging in vivo. *J Neurosci* 32(9):3131-3141.
51. Mani M, Goyal S, Irvine KD, & Shraiman BI (2013) Collective polarization model for gradient sensing via Dachsous-Fat intercellular signaling. *Proc Natl Acad Sci U S A* 110(51):20420-20425.
52. Bialek W, *et al.* (2012) Statistical mechanics for natural flocks of birds. *Proc Natl Acad Sci U S A* 109(13):4786-4791.
53. Cover TM & Thomas JA (1991) *Elements of Information Theory* (John Wiley & Sons, New York).
54. El Gamal AA & Kim Y-H (2011) *Network information theory* (Cambridge University Press, Cambridge ; New York) pp xxviii, 685 p.
55. Lu P, Ewald AJ, Martin GR, & Werb Z (2008) Genetic mosaic analysis reveals FGF receptor 2 function in terminal end buds during mammary gland branching morphogenesis. *Developmental biology* 321(1):77-87.
56. Osborne LC, Lisberger SG, & Bialek W (2005) A sensory source for motor variation. *Nature* 437(7057):412-416.
57. Gregor T, Bialek W, de Ruyter van Steveninck R, Tank D, & Wieschaus E (2005) Diffusion and scalign during early embryonic pattern development. *PNAS* 102:18403-18407.
58. Tsai FC & Meyer T (2012) Ca2+ pulses control local cycles of lamellipodia retraction and adhesion along the front of migrating cells. *Current biology : CB* 22(9):837-842.
59. Haeger A, Wolf K, Zegers MM, & Friedl P (2015) Collective cell migration: guidance principles and hierarchies. *Trends in cell biology*.
60. Huang CP, *et al.* (2009) Engineering microscale cellular niches for three-dimensional multicellular co-cultures. *Lab Chip* 9(12):1740-1748.
61. Ewald AJ (2013) Practical considerations for long-term time-lapse imaging of epithelial morphogenesis in three-dimensional organotypic cultures. *Cold Spring Harbor protocols* 2013(2):100-117.

## Supplementary Materials:

### Cell-cell communication enhances the capacity of cell ensembles to sense shallow gradients during morphogenesis

David Ellison[1], Andrew Mugler[1], Matthew Brennan[1], Sung Hoon Lee, Robert Huebner, Eliah Shamir, Laura A. Woo, Joseph Kim, Patrick Amar, Ilya Nemenman*, Andrew J. Ewald*, Andre Levchenko*

[1] These authors contributed equally to this work
* Corresponding authors

## 1. Measuring bias in organoid branching

To ensure that our determination of response bias is robust to our analysis technique, we measured bias in several different ways (Figure S1), using data for wild-type organoids in the presence of an EGF gradient (Fig. 1D in the main text). Figure S1A shows a histogram of the angles of all branches, irrespective of which organoid the branch comes from. Figure S1B shows a histogram of all organoid angles, where organoid angle is defined as the angle of the vector sum of all branches coming from a given organoid. Thus Fig. S1A is a branch-based histogram, whereas Fig. S1B is an organoid-based histogram. Figures S1A and B demonstrate that both a branch-based and an organoid-based analysis indicate that the response of wild-type organoids is significantly biased in the gradient direction. Figure S1C shows the six different bias measures defined in Fig. 2 of the main text, applied to both the branch-based and the organoid-based data. In all cases, the response is significantly biased with respect to the null value. This demonstrates that the determination of bias is robust to the choice of bias measure.

In general, we find that it does not matter whether we use a branch- or organoid-based measure to determine bias. Therefore, we focus on organoid-based measures for most of the study, since this metric retains the information about the organoids producing the branches, rather than considering branches as completely independent entities. Moreover, in general we also find that the determination of bias is robust to the choice of bias measure (see Figs. S2 and S3 below). Therefore we focus on measure B for most of the study, since it is easy to interpret and to compare with the theory: it is the probability that the vector sum of an organoid's branches points up the gradient, not down the gradient.

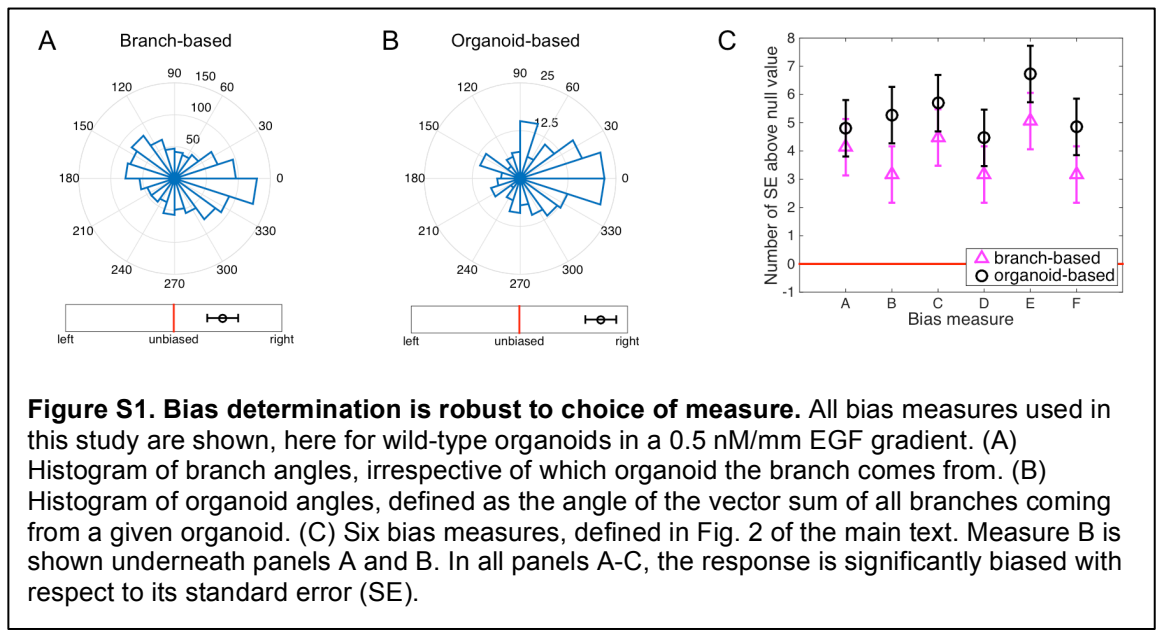

**Figure S1. Bias determination is robust to choice of measure.** All bias measures used in this study are shown, here for wild-type organoids in a 0.5 nM/mm EGF gradient. (A) Histogram of branch angles, irrespective of which organoid the branch comes from. (B) Histogram of organoid angles, defined as the angle of the vector sum of all branches coming from a given organoid. (C) Six bias measures, defined in Fig. 2 of the main text. Measure B is shown underneath panels A and B. In all panels A-C, the response is significantly biased with respect to its standard error (SE).

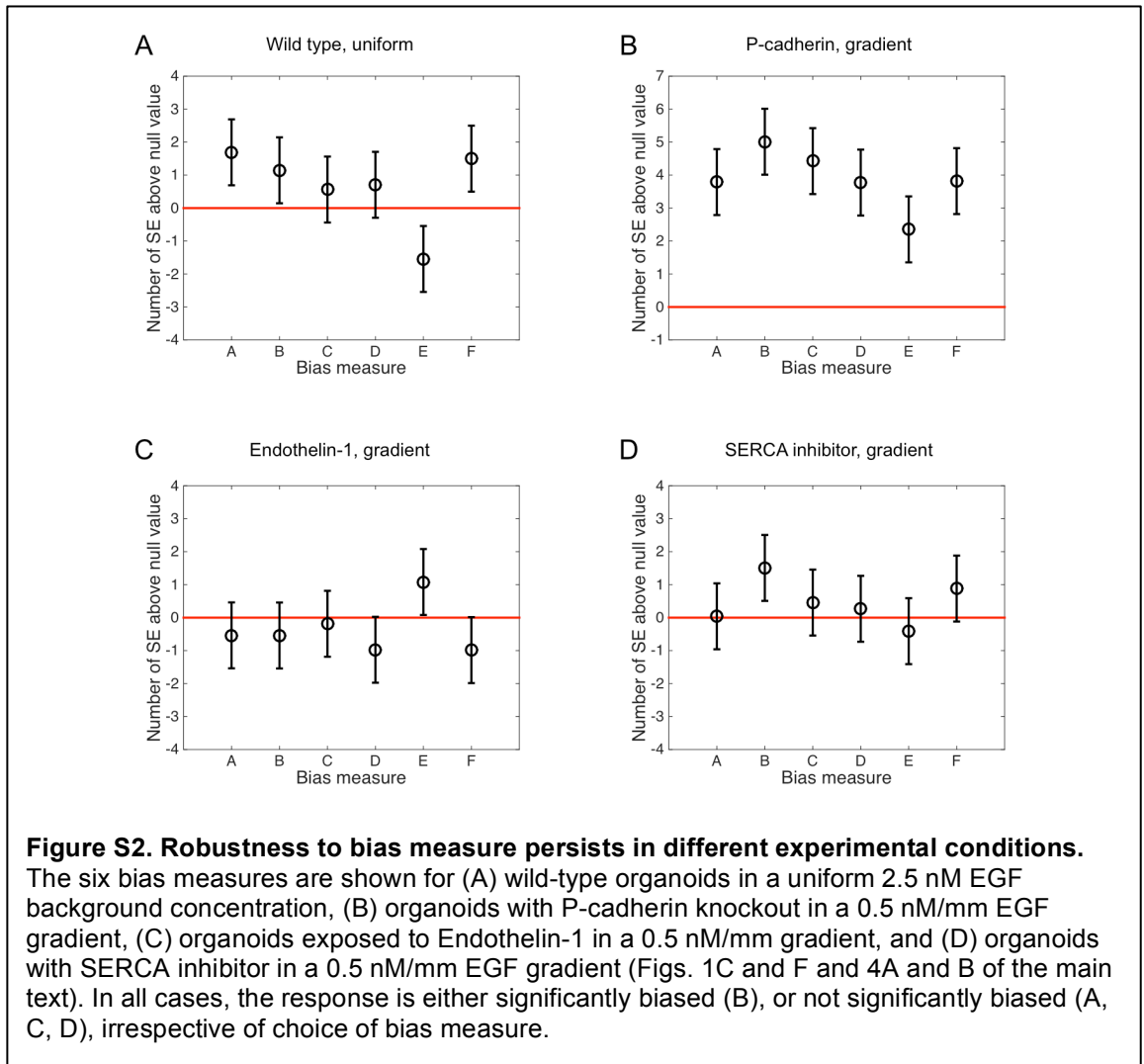

**Figure S2. Robustness to bias measure persists in different experimental conditions.** The six bias measures are shown for (A) wild-type organoids in a uniform 2.5 nM EGF background concentration, (B) organoids with P-cadherin knockout in a 0.5 nM/mm EGF gradient, (C) organoids exposed to Endothelin-1 in a 0.5 nM/mm gradient, and (D) organoids with SERCA inhibitor in a 0.5 nM/mm EGF gradient (Figs. 1C and F and 4A and B of the main text). In all cases, the response is either significantly biased (B), or not significantly biased (A, C, D), irrespective of choice of bias measure.

Figure S2 shows the six bias measures for each of the other experimental conditions considered in the main text. We see in all cases that the presence or absence of bias is robust to the choice of measure.

**2. Measuring bias in single-cell movement**

To ensure that our determination of bias in single-cell movement is also robust to the analysis technique, we subject the single-cell data to a similar multitude of bias measures. For single cells, the analog of a "branch" is the distance the cell migrates over time. Therefore, if more cells migrate to the right than to the left, then the cells exhibit a biased response. Fig. S3 shows the same bias measures computed for organoid branching, but now for single cell migration distances, for the experiment in which the P-cadherin mutation promotes shedding of single cells from the organoid. We compute the bias measures both (i) averaged over all cells, irrespective of the organoid from which cells are shed (Fig. S3A, analogous to the "branch-based" measures in Fig. S1C) and (ii) averaged per organoid, by accounting for the organoid from which the cells are shed (Fig. S3B, analogous to the "organoid-based" measures in Fig. 1C). In both cases, we see that the single-cell movement is not significantly biased, and that the absence of bias is robust to the choice of measure.

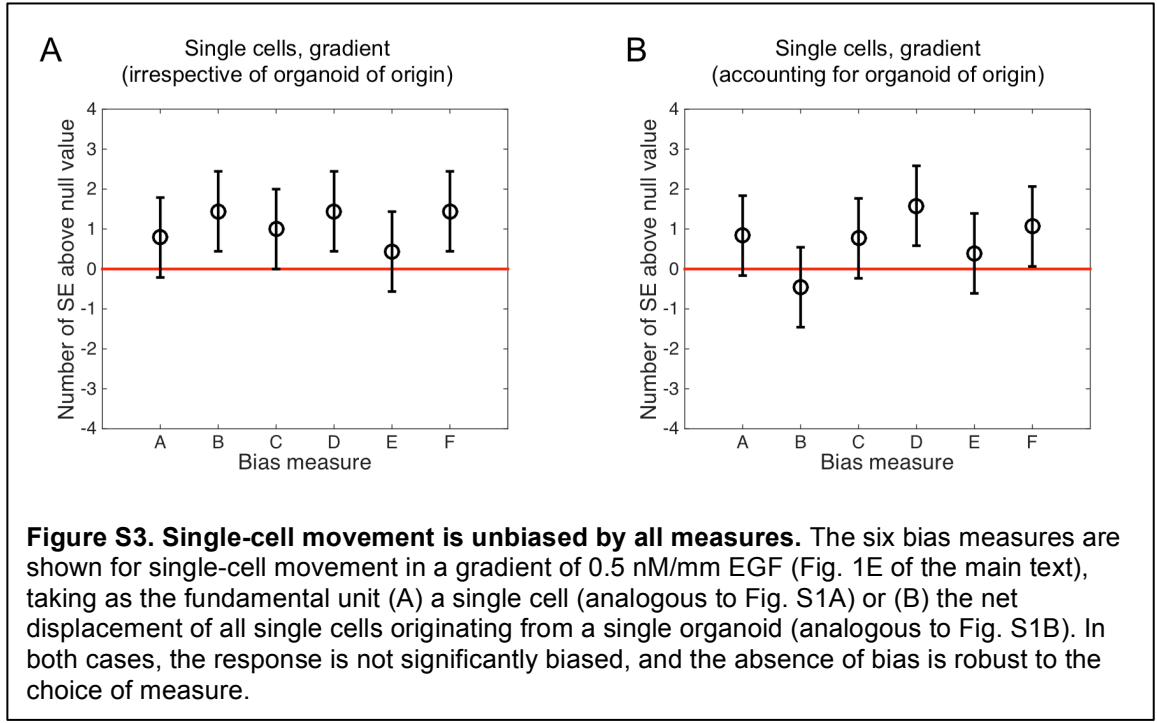

**Figure S3. Single-cell movement is unbiased by all measures.** The six bias measures are shown for single-cell movement in a gradient of 0.5 nM/mm EGF (Fig. 1E of the main text), taking as the fundamental unit (A) a single cell (analogous to Fig. S1A) or (B) the net displacement of all single cells originating from a single organoid (analogous to Fig. S1B). In both cases, the response is not significantly biased, and the absence of bias is robust to the choice of measure.

### 3. Mechanistic model of communicating cells

Here we present the stochastic model of gradient sensing by communicating cells. We consider a one-dimensional chain of $N$ cells parallel to the gradient direction. As in the experiments, the mean EGF signal concentration varies linearly along the direction of the chain as

$$\bar{c}_n = \bar{c}_N - ag(N-n), \tag{1}$$

where $\bar{c}_n$ is the local concentration near the $n$th cell, $a$ is the cell diameter, $g$ is the concentration gradient, and $\bar{c}_N$ is the maximal concentration at the $N$th cell. The observed independence of bias on background concentration (Fig. 3C of the main text) supports an adaptive model of sensing. We therefore choose a minimal adaptive model based on the principle of local excitation and global inhibition (LEGI) [5]. In the $n$th cell, both a local and a global molecular species are degraded at a rate $\mu$ and produced at a rate $\beta$ in proportion to the number of signal molecules in the vicinity, which is roughly $c_n a^3$. Whereas the local species is confined to each cell, the global species is exchanged between neighboring cells at a rate $\gamma$, which provides the communication. Because there is no experimental evidence for receptor saturation (Fig. 3C of the main text), we confine ourselves to the linear response regime, in which the dynamics of the local and global species satisfy the stochastic equations

$$\frac{dx_n}{dt} = \beta(c_n a^3) - \mu x_n + \eta_n, \tag{2}$$

$$\frac{dy_n}{dt} = \beta(c_n a^3) - \mu y_n + \gamma(y_{n-1} + y_{n+1} - 2y_n) + \xi_n$$
$$= \beta(c_n a^3) - \mu \sum_{n'=1}^{N} M_{nn'} y_{n'} + \xi_n, \tag{3}$$

where

$$M_{nn'} \equiv (1 + 2\gamma/\mu)\delta_{nn'} - (\gamma/\mu)(\delta_{n',n-1} + \delta_{n',n+1}) \tag{4}$$

is the tridiagonal matrix governing degradation and exchange. Here $x_n$ and $y_n$ are the molecule numbers of the local and global species, respectively, and the terms $\eta_n$ and $\xi_n$ are the intrinsic Langevin noise terms with zero mean and covariances

$$\langle \eta_n(t)\eta_{n'}(t')\rangle = (\beta \bar{c}_n a^3 + \mu \bar{x}_n)\delta_{nn'}\delta(t-t'), \tag{5}$$
$$\langle \xi_n(t)\xi_{n'}(t')\rangle = [(\beta \bar{c}_n a^3 + \mu \bar{y}_n + \gamma \bar{y}_{n-1} + \gamma \bar{y}_{n+1} + 2\gamma \bar{y}_n)\delta_{nn'}$$
$$- (\gamma \bar{y}_n + \gamma \bar{y}_{n-1})\delta_{n',n-1} - (\gamma \bar{y}_n + \gamma \bar{y}_{n+1})\delta_{n',n+1}]\delta(t-t'). \tag{6}$$

Equation 5 and the first line of Eq. 6 contain the Poisson noise corresponding to each reaction, while the second line of Eq. 6 contains the anti-correlations between neighboring cells introduced by the exchange. Equations 4 and 6 are modified at the edges $n = \{1, N\}$ to include exchange with just one neighboring cell.

In the LEGI framework, the local species excites a downstream species, while the global species inhibits it. In the limit of shallow gradients, the relative noise in the excitation level of this downstream species is equivalent to that in the difference $\Delta_n = x_n - y_n$ between local and global species' molecule numbers. To see this, we recall from Ref. [5] that, in the LEGI model, the excitation level $r$ depends on the ratio of activator $x$ to inhibitor $y$ as

$$r = \frac{x/y}{x/y + z}, \tag{7}$$

where $z$ is a constant. At equal activation and inhibition, $x = y$, the excitation level is $r_0 = 1/(1+z)$. Defining $s \equiv r - r_0$ as the deviation from this level, Eq. 7 can be written in terms of $\Delta = x - y$ and $y$ as

$$s = \frac{z}{(1+z)^2} \frac{\Delta}{y}, \tag{8}$$

where, for shallow gradients, we have assumed that the quantity $\Delta/y$ is small. Small fluctuations among $s$, $\Delta$, and $y$ are therefore related as

$$\delta s = \frac{z}{(1+z)^2}\left(\frac{\delta\Delta}{\bar{y}} - \frac{\bar{\Delta}}{\bar{y}^2}\delta y\right), \tag{9}$$

or equivalently,

$$\left(\frac{\delta s}{\bar{s}}\right)^2 = \left(\frac{\delta\Delta}{\bar{\Delta}}\right)^2\left(1 - \frac{\bar{\Delta}}{\bar{y}}\frac{\delta y}{\delta\Delta}\right)^2 \approx \left(\frac{\delta\Delta}{\bar{\Delta}}\right)^2, \tag{10}$$

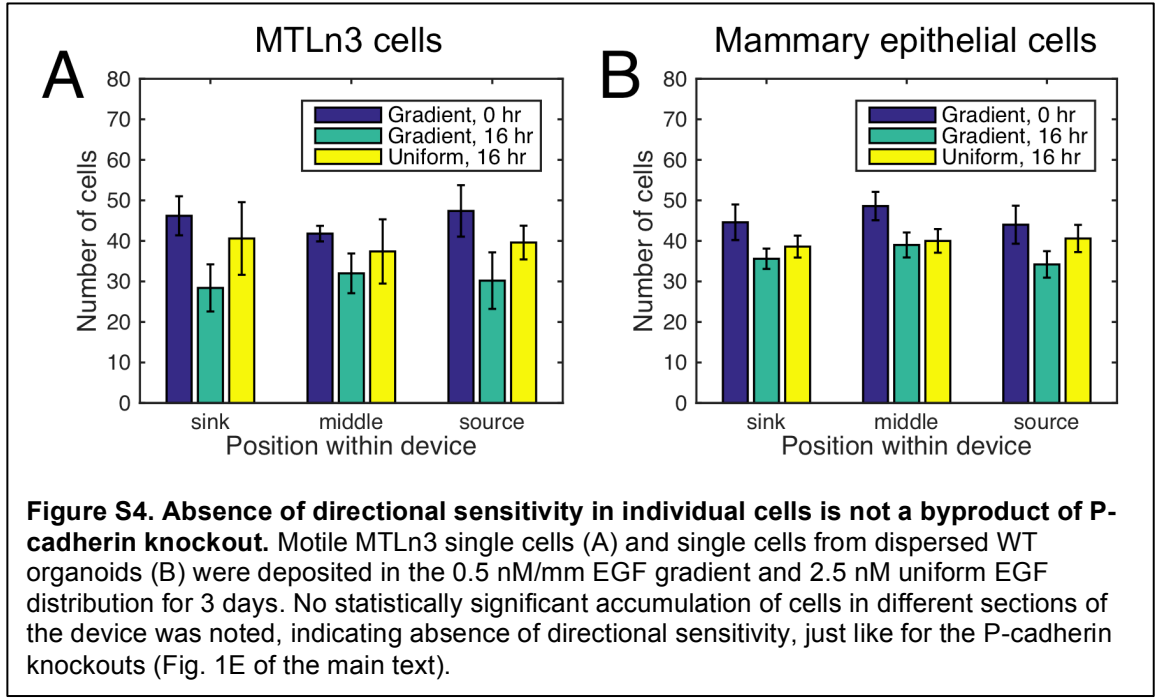

**Figure S4. Absence of directional sensitivity in individual cells is not a byproduct of P-cadherin knockout.** Motile MTLn3 single cells (A) and single cells from dispersed WT organoids (B) were deposited in the 0.5 nM/mm EGF gradient and 2.5 nM uniform EGF distribution for 3 days. No statistically significant accumulation of cells in different sections of the device was noted, indicating absence of directional sensitivity, just like for the P-cadherin knockouts (Fig. 1E of the main text).

where the last step once again assumes $\bar{\Delta}/\bar{y}$ is small. Thus we see that relative fluctuations in $s$ are equivalent to those in $\Delta$. We therefore take $\Delta$ as our readout variable, focusing in particular on $\Delta_N$, the molecule number difference in the cell furthest up the gradient, since this cell initiates the morphological branching observed in the experiment.

**4. Absence of directional sensitivity in single cells**

While Fig. 1E showed the absence of directional sensitivity in individual cells, it remains possible that this insensitivity is a result of the P-cadherin knockout. To alleviate this possibility, we deposited individual cells from the MTLn3 mammary epithelial cell line [6], as well as individual cells from dispersed WT organoids into the experimental device for 3 days. Over this time, the cells can move over distances comparable to those determined organoid experiments. Directionally biased motility would result in enrichment of cells in different device zones (source of EGF, middle of the device, and sink of EGF). As seen in Supp. Fig. S5, no enrichment is observed, indicating that the absence of directional sensitivity in individual cells is not a byproduct of the P-cadherin knockout.

**5. Instantaneous vs. temporally-integrated gradient sensing**

Since the foundational publication of Berg and Purcell [1], most work on molecular sensing has considered the setup where a sensor integrates the signal over a certain time $t$, much larger than the typical turnover time of the ligand molecules, which is controlled by diffusion. As the diffusion brings new molecules to the vicinity of the sensor, fluctuations are averaged out, resulting in a typical $\sim 1/\sqrt{t}$ decrease of the sensory error. Analyses of gradient sensing not considering [2] and considering communication [7] among the neighboring cells have also revealed similar time dependence due to temporal integration. In contrast, Figure S4 shows the organoids do not exhibit an increase in sensory precision with time between 1 and 3 days of the experiment duration. This suggests that the integration (or memory) time in this system is smaller than the typical diffusive turnover time. As a consistency check, we point out that the diffusion coefficient of EGF in extracellular space is about 50 um$^2$/s [8]. Thus a typical diffusion time across a 300 um organoid would be (300 um)$^2$/(50 um$^2$/s) = 30 min, so many biochemical signaling reactions – and integration scales defined by them – are faster (see [7] for a more careful analysis of time scales relevant for collective gradient sensing). Therefore, in what follows, we consider that $\Delta$, an instantaneous steady-state, rather than time-averaged, difference of the local and the global messenger species, is the readout of our model most relevant for the experiments. At the same time, we refer the reader to the companion article, Ref. [7], where a full analysis with temporal integration is presented. The integration does not change the qualitative picture developed here (existence of a finite gradient sensing unit), but provides somewhat different values for the dependence of the sensory limits on the system parameters.

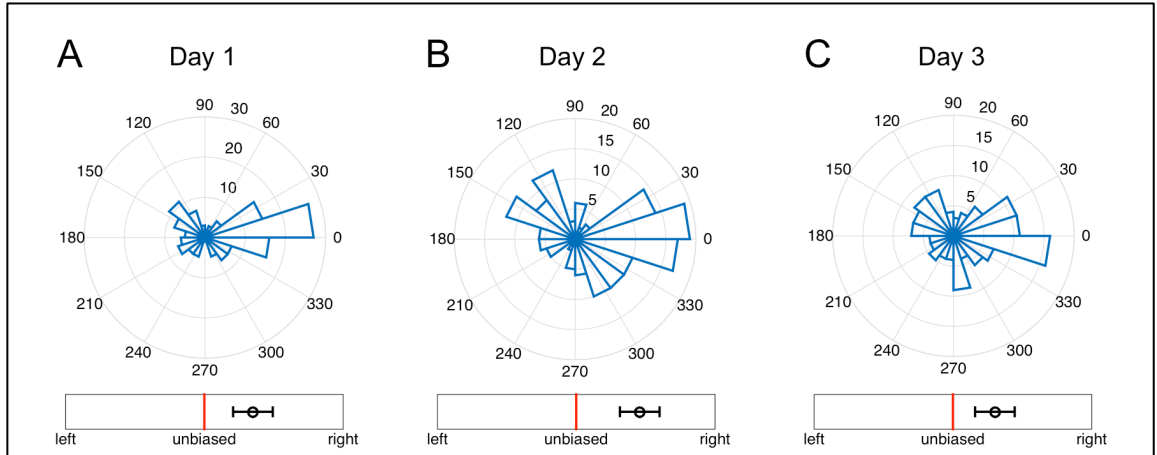

**Figure S5. Temporal stability of gradient sensing.** Angular histograms of branch directions for new branches appearing during (A) day 1 (total 145 branches), (B) day 2 (total 157 branches), and (C) day 3 (total 132 branches) of continual exposure of organoids to a 0.5 nM/mm EGF gradient. Branch angles are plotted irrespective of the organoid from which each branch originates. The gradient of EGF is in the 0° direction. Plots are non-cumulative, in that B shows only branches that form during day 2 (and not day 1), and C shows only branches that form during day 3 (and not days 1 and 2). The number of branches changes since the organoids change their morphology with time, but there is clearly no evidence of an improving accuracy of branch formation angle, and hence no evidence of temporal integration. Bias is measured by measure B from Fig. 2 in the main text (but for individual branches, as in Fig. S1).

### 6. Mean and variance of the readout variable

The mean and variance of the readout variable are

$$\bar{\Delta}_N = \bar{x}_N - \bar{y}_N, \tag{11}$$

$$(\delta \Delta_N)^2 = (\delta x_N)^2 + (\delta y_N)^2 - 2 \, \text{cov}(x_N, y_N), \tag{12}$$

where $\text{cov}(x_N, y_N) = \langle (x_N - \bar{x}_N)(y_N - \bar{y}_N) \rangle$ is the covariance. These expressions in turn depend on the mean and variance of $x_N$ and $y_N$, which we now calculate from Eqs. 2 and 3 in steady state. The mean of $x_N$ follows straightforwardly from Eq. 2,

$$\bar{x}_N = G\bar{c}_N a^3, \tag{13}$$

where the term $G \equiv \beta/\mu$ describes the factor by which the number of local species molecules is amplified beyond the number of detected signal molecules. Similarly, the mean of $y_N$ follows from Eq. 3,

$$\bar{y}_N = G \sum_{n=0}^{N-1} K_n \bar{c}_{N-n} a^3, \tag{14}$$

where $K_n \equiv M^{-1}_{N,N-n}$. We see that, due to the communication, the global species number in the edge cell is a weighted sum of the signal measurements made by all the other cells. The weighting is determined by $K_n$, which we call the communication kernel and discuss in detail in the next section.

The variance of $x_N$ is easiest to derive in Fourier space. We first consider the fluctuations $\delta x_N = x_N - \bar{x}_N$ and $\delta c_N = c_N - \bar{c}_N$, in terms of which Eq. 2 reads

$$\frac{d\delta x_N}{dt} = \beta \delta c_N a^3 - \mu \delta x_N + \eta_N. \tag{15}$$

Fourier transforming and rearranging obtains

$$\tilde{\delta x}_N = \frac{\beta \tilde{\delta c}_N a^3 + \tilde{\eta}_N}{\mu - i\omega}. \tag{16}$$

Since we are interested in the instantaneous readouts only, the variance is then the integral over all frequencies of the power spectrum $S_x(\omega) = \langle \tilde{\delta x}_N^* \tilde{\delta x}_N \rangle$,

$$(\delta x_N)^2 = \int_{-\infty}^{\infty} \frac{d\omega}{2\pi} S_x(\omega) = \beta^2 \int_{-\infty}^{\infty} \frac{d\omega}{2\pi} \frac{S_c(\omega)a^6}{\mu^2 + \omega^2} + \int_{-\infty}^{\infty} \frac{d\omega}{2\pi} \frac{\langle \tilde{\eta}_N^* \tilde{\eta}_N \rangle}{\mu^2 + \omega^2}, \tag{17}$$

where the cross terms vanish because signal fluctuations are not cross-correlated with local species fluctuations. The noise spectrum follows from Eq. 5, $\langle \tilde{\eta}_N^* \tilde{\eta}_N \rangle = \beta \bar{c}_N + \mu \bar{x}_N = 2\mu \bar{x}_N$, upon which the second term in Eq. 17 integrates to $\bar{x}_N$. The first term in Eq. 17 depends on the power spectrum of signal fluctuations, which for a Poisson process with timescale $\tau$ reads $S_c(\omega)a^6 = 2\tau \bar{c}_N a^3/[1 + (\omega\tau)^2]$. We are considering instantaneous readouts, which is equivalent to the diffusion of EGF being slow, i.e., $\tau \to \infty$ and $S_c(\omega)a^6 \to 2\pi\delta(\omega)\bar{c}_N a^3$. This is the same as assuming that the number of signal molecules is Poisson-distributed but fixed in time. Thus Eq. 17 becomes

$$(\delta x_N)^2 = G^2 \bar{c}_N a^3 + \bar{x}_N. \tag{18}$$

The first term is the extrinsic noise. It arises from fluctuations in the signal molecule number. Since these fluctuations are Poissonian, the variance of the signal molecule number equals its mean $\bar{c}_N a^3$. Then, as these fluctuations are propagated to the local species, they are amplified by the gain $G^2$. The second term is the intrinsic noise. The intrinsic noise arises from fluctuations in the local species number itself. These fluctuations are also Poissonian, and thus the variance equals the mean $\bar{x}_N$.

We follow the same procedure to find the variance of $y_N$. The result is

$$(\delta y_N)^2 = G^2 \sum_{n=0}^{N-1} K_n^2 \bar{c}_{N-n} a^3 + \bar{y}_N. \tag{19}$$

The extrinsic noise (first term) once again scales with the gain $G^2$. It depends on the same kernel $K_n$ that determines the mean, which reflects the fact that, as seen in Eq. 15, upstream fluctuations propagate through linear systems in the same way as the signals themselves [9]. The intrinsic noise (second term) is once again equal to the mean $\bar{y}_N$, which is a necessary consequence of the fact that Eq. 3 is an open system whose reaction rates are linear in the species numbers [10].

Finally, we apply the same technique to find the covariance, which is the integral over all frequencies of the cross-spectrum $\langle \delta \tilde{x}_N^* \delta \tilde{y}_N \rangle$. The result is

$$\text{cov}(x_N, y_N) = G^2 K_0 \bar{c}_N a^3. \tag{20}$$

This expression has a straightforward interpretation: it is the product of two extrinsic standard deviations. The first is the square root of the extrinsic noise in the local species, $\sigma_x \equiv \sqrt{G^2 \bar{c}_N a^3}$. The second is the square root of the extrinsic noise in the global species, but only the component affecting the $N$th cell, $\sigma_y \equiv \sqrt{G^2 K_0^2 \bar{c}_N a^3}$. The reason that only extrinsic noise enters is because $x_N$ and $y_N$ only co-vary due to fluctuations in the extrinsic signal. The reason that only the $N$th component of the global noise contributes is because the local species is not communicated, and thus any effect on $y_N$ due to other cells cannot co-vary with $x_N$. Finally, the reason that the covariance takes the form of a product of standard deviations is because $x_N$ and $y_N$ depend identically on the signal (Eqs. 2 and 3), and therefore the correlation coefficient corresponding to extrinsic fluctuations $\text{cov}_{\text{extrinsic}}(x_N, y_N)/(\sigma_{x_N} \sigma_{y_N})$ is equal to one.

From the mean, variance, and covariance of $x_N$ and $y_N$, the mean and variance of the readout variable follow via Eqs. 11 and 12. The only thing that remains is to solve for the communication kernel $K_n$, which we describe next.

### 7. Communication kernel

The communication kernel $K_n \equiv M_{N,N-n}^{-1}$ is found by inverting the tridiagonal matrix $M_{nn'}$. First we derive the inverse, and then we present an approximation of $K_n$ in the limit of strong communication and many cells.

Defining $\rho \equiv \mu/\gamma$, the diagonal ($u$), superdiagonal ($v$), and subdiagonal ($w$) terms of $\rho M_{nn'}$ (Eq. 4) are

$$u_n = \begin{cases} \rho + 1 & n = 1, \\ \rho + 2 & 2 \leq n \leq N-1, \\ \rho + 1 & n = N, \end{cases}$$
$$v_n = -1 \qquad 1 \leq n \leq N-1,$$
$$w_n = -1 \qquad 1 \leq n \leq N-1. \tag{21}$$

The inverse of any tridiagonal matrix can be calculated by recursion [11, 12],

$$M_{nn'}^{-1} = \rho \begin{cases} (-1)^{n+n'} v_n \ldots v_{n'-1} \theta_{n-1} \phi_{n'+1}/\theta_N & n \leq n', \\ (-1)^{n+n'} w_{n'} \ldots w_{n-1} \theta_{n'-1} \phi_{n+1}/\theta_N & n > n', \end{cases} \tag{22}$$

where $\theta_n$ and $\phi_n$ satisfy

$$\begin{aligned} \theta_0 &= 1, & \theta_1 &= u_1, & \theta_n &= u_n \theta_{n-1} - v_{n-1} w_{n-1} \theta_{n-2} & 2 \leq n \leq N, \\ \phi_{N+1} &= 1, & \phi_N &= u_N, & \phi_n &= u_n \phi_{n+1} - v_n w_n \phi_{n+2} & N-1 \geq n \geq 1. \end{aligned} \tag{23}$$

Since both $v_n$ and $w_n$ are constant and equal to $-1$, Eq. 22 simplifies to

$$M_{nn'}^{-1} = \rho \begin{cases} \theta_{n-1} \phi_{n'+1}/\theta_N & n \leq n', \\ \theta_{n'-1} \phi_{n+1}/\theta_N & n > n'. \end{cases} \tag{24}$$

From Eq. 24 we can also deduce that the inverse is symmetric. We write the first few terms of $\theta_n$ and notice the pattern,

$$\begin{aligned} \theta_0 &= 1, \\ \theta_1 &= \rho + 1, \\ \theta_2 &= (\rho+2)(\rho+1) - 1 = \rho^2 + 3\rho + 1, \\ \theta_3 &= (\rho+2)[(\rho+2)(\rho+1) - 1] - (\rho+1) = \rho^3 + 5\rho^2 + 6\rho + 1, \\ &\vdots \\ \theta_n &= \sum_{j=0}^{n} \binom{n+j}{2j} \rho^j \quad 0 \leq n \leq N-1. \end{aligned} \tag{25}$$

The last term $\theta_N$ does not conform to the pattern because $u_N$ is different from its previous terms, so we calculate $\theta_N$ explicitly from $\theta_{N-1}$ and $\theta_{N-2}$ and simplify,

$$\theta_N = \sum_{j=1}^{N} \binom{N+j-1}{2j-1} \rho^j. \tag{26}$$

Then, since $v_n$ and $w_n$ are constants and $u_n = u_{N-n+1}$, we notice from Eq. 23 that

$$\phi_n = \theta_{N-n+1}. \tag{27}$$

Inserting Eqs. 25-27 into Eq. 24 and simplifying, and recalling that the inverse is symmetric, we

arrive at the expression

$$M^{-1}_{nn'} = M^{-1}_{n'n} = \frac{\sum_{j=0}^{n-1}\binom{n-1+j}{2j}\rho^j \sum_{k=0}^{N-n'}\binom{N-n'+k}{2k}\rho^k}{\sum_{\ell=0}^{N-1}\binom{N+\ell}{2\ell+1}\rho^\ell} \quad n \leq n' \tag{28}$$

for the inverse. The communication kernel is a particular case,

$$K_n \equiv M^{-1}_{N,N-n} = \frac{\sum_{j=0}^{N-n-1}\binom{N-n-1+j}{2j}\rho^j}{\sum_{\ell=0}^{N-1}\binom{N+l}{2l+1}\rho^\ell}. \tag{29}$$

The communication kernel is normalized, $\sum_{n=1}^{N} K_n = 1$, which is consistent with its interpretation as a weighting function.

Now we show that in the limit of strong communication and a large number of cells, the communication kernel can be approximated by an exponential distribution. Since all dependence on $n$ occurs in the numerator of Eq. 29, we approximate the numerator only, and then we set the denominator using the fact that $K_n$ is normalized. The approximation of the numerator follows two steps. First, the factorials in the choose function are written using the Stirling approximation. Second, the sum is simplified using the saddle point approximation.

We expect $K_n$ to have the strongest support at the edge cell and nearby cells, i.e. for small values of $n$. Therefore, applying the Stirling approximation to the numerator of Eq. 29 is valid in the limit

$$N \gg j^* \gg 1, \tag{30}$$

where $j^*$ is the value at which the summand peaks. We will see below that this condition is satisfied in the limit of strong communication and many cells.

Ignoring the denominator, we write the exchange kernel as $K_n \propto \sum_j e^{-g_j}$, where

$$\begin{aligned} g_j &\equiv -\log\left[\binom{N-n-1+j}{2j}\rho^j\right] \\ &= -\log[(N-n-1+j)!] + \log[(N-n-1-j)!] + \log[(2j)!] - j\log\rho. \end{aligned} \tag{31}$$

Applying the Stirling approximation $\log(x!) = (x+1/2)\log x - x + (1/2)\log(2\pi)$ yields

$$\begin{aligned} g_j = &-\left(N-n+j-\frac{1}{2}\right)\log(N-n+j-1) + \left(N-n-j-\frac{1}{2}\right)\log(N-n-j-1) \\ &+ \left(2j+\frac{1}{2}\right)\log(2j) - j\log\rho + \frac{1}{2}\log 2\pi. \end{aligned} \tag{32}$$

We now apply the saddle point approximation, which means we approximate $j$ as continuous and expand $g_j$ to second order around its minimum value, permitting the evaluation of a Gaussian integral,

$$K_n \propto \sum_j e^{-g_j} \approx \int dj\, \exp\left[-g_* - \frac{1}{2}g''_*(j-j_*)^2\right] = e^{-g_*}\sqrt{\frac{\pi}{g''_*}}. \tag{33}$$

Here $j^*$ is the value at which the minimum $g_*$ occurs and at which the second derivative $g_*''$ is evaluated. It is found by setting to zero the first derivative of Eq. 32,

$$g_j' = -\log(N-n+j-1) - \log(N-n-j-1) + 2\log(2j) - \log\rho$$
$$- \frac{1}{2(N-n+j-1)} - \frac{1}{2(N-n-j-1)} + \frac{1}{2j}. \tag{34}$$

Ignoring the last three terms because their denominators are precisely the three quantities we have assumed are large, we solve $g_j' = 0$ to find

$$j^* = \psi(N-n-1), \tag{35}$$

where $\psi \equiv \sqrt{\rho/(\rho+4)}$. Eq. 35 shows that $j^* \sim \psi N$, which means Eq. 30 can be written

$$1 \gg \psi \gg 1/N. \tag{36}$$

The left condition in Eq. 36 requires that $\psi$ is small. This is satisfied in the strong communication limit $\gamma \gg \mu$, since then $\psi \approx \sqrt{\rho}/2 = \sqrt{\mu/\gamma}/2 \ll 1$. The right condition in Eq. 36 requires that $N$ is large (there are many cells), such that the kernel falls to nearly zero still within the organoid.

Inserting Eq. 35 value into Eq. 32 yields

$$g_* = -\left(N-n-\frac{1}{2}\right)\log\left(\frac{1+\psi}{1-\psi}\right) + \frac{1}{2}\log[4\pi\psi(n-1)]. \tag{37}$$

Then differentiating Eq. 34, once again ignoring the last three terms,

$$g_j'' = \frac{-1}{N-n+j-1} + \frac{1}{N-n-j-1} + \frac{2}{j}, \tag{38}$$

and inserting Eq. 35 yields

$$g_*'' = \frac{-1}{(1+\psi)(N-n-1)} + \frac{1}{(1-\psi)(N-n-1)} + \frac{2}{\psi(N-n-1)}$$
$$= \frac{2}{\psi(1-\psi^2)(N-n-1)}. \tag{39}$$

Now we evaluate the saddle point result (Eq. 33),

$$K_n \propto \exp\left[\left(N-n-\frac{1}{2}\right)\log\left(\frac{1+\psi}{1-\psi}\right)\right]\sqrt{\frac{1}{4\pi\psi(N-n-1)}}\sqrt{\frac{\pi\psi(1-\psi^2)(N-n-1)}{2}}$$
$$\propto e^{-n/n_0}, \tag{40}$$

where in the second step we drop all $n$-independent prefactors and define $n_0 \equiv 1/\log[(1+\psi)/(1-\psi)]$. We recover the proper prefactor by enforcing normalization,

$1 = \sum_n K_n \approx \int dn \, K_n$,

$$K_n \approx \frac{1}{n_0} e^{-n/n_0}, \tag{41}$$

and we see that the kernel falls off exponentially with the number of cells $n$ from the edge cell.

The kernel length scale $n_0$ can be simplified in the strong communication limit, in which $\psi \approx \sqrt{\rho}/2$ is small,

$$n_0 \approx \frac{1}{\log(1+2\psi)} \approx \frac{1}{2\psi} \approx \frac{1}{\sqrt{\rho}} = \sqrt{\frac{\gamma}{\mu}}. \tag{42}$$

We see that the length scale is the square root of the ratio of a diffusion term ($\gamma$) to a degradation term ($\mu$). This is the same form as the length scale of morphogen profiles that are set up by diffusion and degradation, which, like the communication kernel, are exponential in shape [13].

## 8. Fundamental limit to the precision of instantaneous gradient sensing with communication

We now complete our calculation of the relative noise in the readout variable $\Delta_N$. In the strong communication and many cells limit, the sums in Eqs. 14 and 19 can be approximated as integrals over all positive $n$ that are then easily evaluated using the exponential form of the kernel (Eq. 41) due to the linearity of $\bar{c}_n$ in $n$ (Eq. 1). We insert the results, along with Eqs. 13, 18, and 20, into Eqs. 11 and 12 to obtain

$$\bar{\Delta}_N = \bar{x}_N - \bar{y}_N = G\bar{c}_N a^3 - G\bar{c}_{N-n_0}a^3 = G(n_0 ag)a^3, \tag{43}$$

$$(\delta\Delta_N)^2 = \left(G^2 \bar{c}_N a^3 + \bar{x}_N\right) + \left(G^2 \frac{\bar{c}_{N-n_0/2}}{2n_0} a^3 + \bar{y}_N\right) - 2\left(G^2 \frac{\bar{c}_N}{n_0} a^3\right), \tag{44}$$

From Eqs. 43 and 44 we obtain the relative noise

$$\left(\frac{\delta\Delta_N}{\bar{\Delta}_N}\right)^2 = \frac{1}{a^3(n_0 ag)^2}\left[\left(\bar{c}_N + \frac{\bar{c}_{N-n_0/2}}{2n_0} - 2\frac{\bar{c}_N}{n_0}\right) + \frac{1}{G^2}\left(\bar{c}_N + \bar{c}_{N-n_0}\right)\right]. \tag{45}$$

Eq. 45 gives the relative uncertainty in the system's estimate of the gradient via its readout $\Delta_N$, in the limit of many cells. In the brackets, the first term in parentheses arises due to the extrinsic noise. The second term in parentheses arises due to the intrinsic noise. The extrinsic and intrinsic terms have a similar structure, and in general as a function of $N$ they will have a similar shape, because they both arise from the same kernel (Eq. 29). The intrinsic term reflects the counting noise from the finite number of internal communicating molecules. The extrinsic noise reflects the imperfect averaging performed by the global molecular species, since it has a finite communication length scale.

In principle, the intrinsic noise can be made arbitrarily small by producing more local and global species molecules, which is equivalent to increasing the gain $G$. Moreover, we observe that in the extrinsic noise, the second and third terms are smaller than the first term by a factor of $n_0$. This is because these terms, which involve the global species, benefit from measurements of the external signal across roughly $n_0$ cells due to the communication. These terms are therefore small relative to the first in the strong communication limit. We are then left with

$$\left(\frac{\delta\Delta_N}{\bar{\Delta}_N}\right)^2 = \frac{\bar{c}_N}{a^3(n_0 ag)^2}. \tag{46}$$

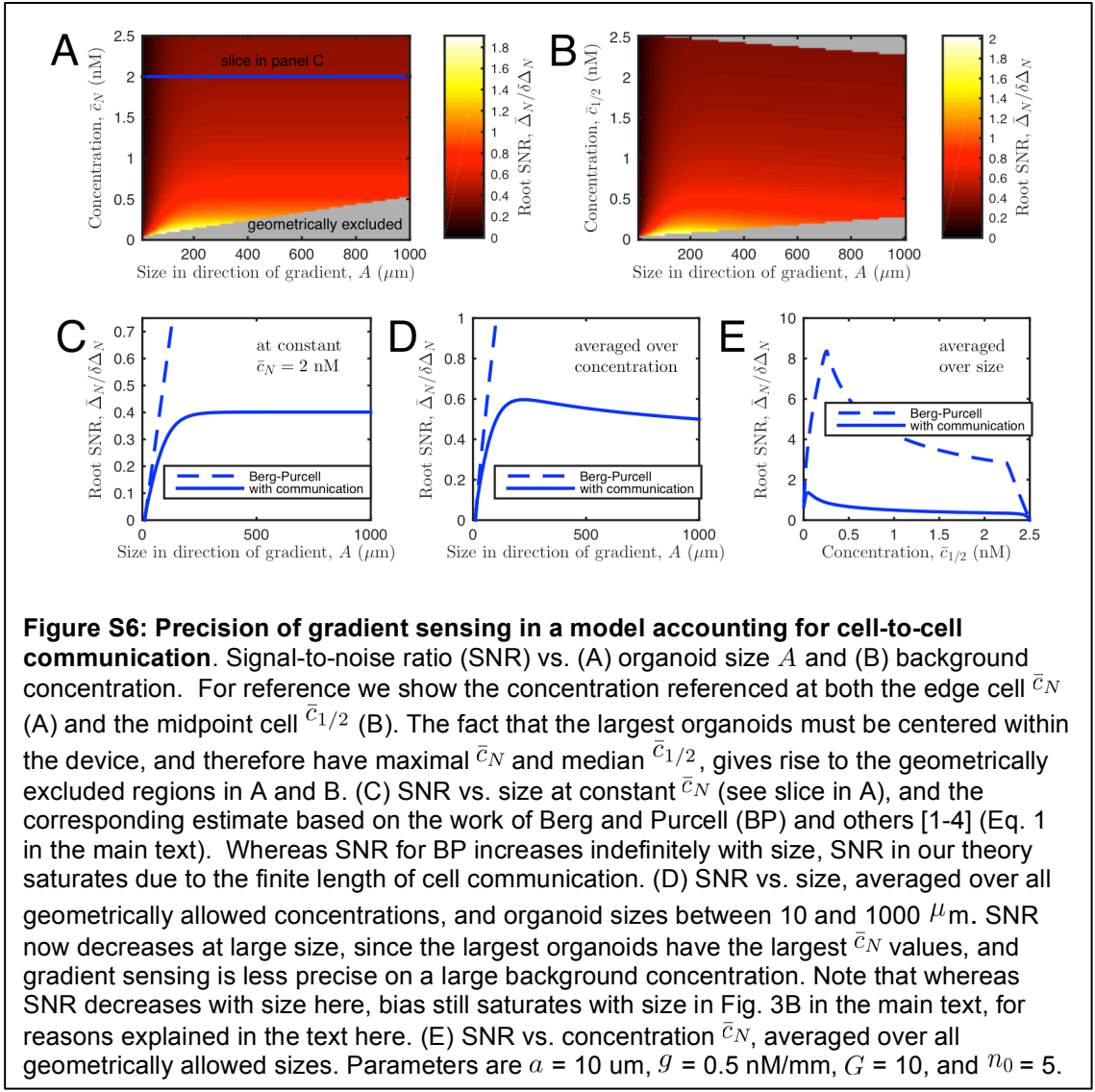

**Figure S6: Precision of gradient sensing in a model accounting for cell-to-cell communication**. Signal-to-noise ratio (SNR) vs. (A) organoid size $A$ and (B) background concentration. For reference we show the concentration referenced at both the edge cell $\bar{c}_N$ (A) and the midpoint cell $\bar{c}_{1/2}$ (B). The fact that the largest organoids must be centered within the device, and therefore have maximal $\bar{c}_N$ and median $\bar{c}_{1/2}$, gives rise to the geometrically excluded regions in A and B. (C) SNR vs. size at constant $\bar{c}_N$ (see slice in A), and the corresponding estimate based on the work of Berg and Purcell (BP) and others [1-4] (Eq. 1 in the main text). Whereas SNR for BP increases indefinitely with size, SNR in our theory saturates due to the finite length of cell communication. (D) SNR vs. size, averaged over all geometrically allowed concentrations, and organoid sizes between 10 and 1000 $\mu$m. SNR now decreases at large size, since the largest organoids have the largest $\bar{c}_N$ values, and gradient sensing is less precise on a large background concentration. Note that whereas SNR decreases with size here, bias still saturates with size in Fig. 3B in the main text, for reasons explained in the text here. (E) SNR vs. concentration $\bar{c}_N$, averaged over all geometrically allowed sizes. Parameters are $a$ = 10 um, $g$ = 0.5 nM/mm, $G$ = 10, and $n_0$ = 5.

This is the central result of this section. Eq. 46 is the fundamental limit to the precision of instantaneous (not temporally averaged) gradient sensing via a LEGI-style adaptive, communicating system. Unlike for a system with temporal integration [7], Eq. 46 does not depend on the measurement time and depends on the spatial averaging scale as $\sim 1/n_0^2$.

Figure S6 shows the values of $(\text{SNR}_N)^{1/2} = \bar{\Delta}_N/\delta\Delta_N$, from Eqs. 11, 12 with the limiting values, or the fundamental limits, given by Eq. 45. In particular, Fig. S6D and E are the analogs of Fig. 3B and C in the main text, except that Fig. 3 plots the estimate of the organoid bias, $P(\Delta_N > \Delta_1)$, which is easily obtained from $\text{SNR}_n$. Note that in Fig. S6D, $\text{SNR}_N$ decreases at large $N$ because large organoids push the $N$th cell to higher concentrations, where gradient sensing is less precise. In contrast, in Fig. 3B in the main text, the bias $P(\Delta_N > \Delta_1)$ saturates, for two reasons: (i) bias derives from the both $\text{SNR}_1$ and $\text{SNR}_N$, which are pushed to opposite concentration regimes for large organoids, and (ii) Fig. 3 also includes additive downstream noise,

| Reaction | Rate |
|---|---|
| $A_n^* + R_n \to A_n^* R_n$ | 5e-2 |
| $A_n^* R_n \to A_n^* + R_n$ | 5e-4 |
| $A_n^* R_n \to A_n^* + R_n^*$ | 1e-3 |
| $I_n^* + R_n^* \to I_n^* R_n^*$ | 5e-2 |
| $I_n^* R_n^* \to I_n^* + R_n^*$ | 5e-4 |
| $I_n^* R_n^* \to I_n^* + R_n$ | 1e-3 |
| $S_n + A_n \to S_n A_n$ | 5e-2 |
| $S_n A_n \to S_n + A_n$ | 5e-4 |
| $S_n A_n \to S_n + A_n^*$ | 1e-3 |
| $A_n^* \to A_n$ | 1e-4 |
| $I_n + S_n \to I_n S_n$ | 5e-2 |
| $S_n I_n \to S_n + I_n$ | 5e-4 |
| $S_n I_n \to S_n + I_n^*$ | 1e-3 |
| $I_n^* \to I_n$ | 1e-4 |
| $I_n^* \to I_{n+1}^*$ | 1e-2 |
| $I_n^* \to I_{n-1}^*$ | 1e-2 |
| $I_n \to I_{n+1}$ | 1e-2 |
| $I_n \to I_{n-1}$ | 1e-2 |

**Table S1: Simulation parameters** used in the spatially-extended Gillespie simulations with low saturation.

which is independent of both size and concentration, and thus tends to flatten out dependencies.

## 9. Spatially resolved Gillespie stochastic simulations to explore modification of fundamental limits to the precision of instantaneous gradient sensing under violation of linearity assumptions

Our theory above made two linearity assumptions. First, we assumed that receptors are not saturated at high ligand concentrations, allowing us to treat the production rate of messenger molecules as a linear function of the position. Second, we assumed that the readout is the difference of the local and the diffusive messenger. In more conventional analysis of LEGI models, the readout is the concentration a response molecule *R*, positively modified by the activator *A* and negatively modified by the inhibitor, *I* [5, 14, 15]. To verify how our findings for the fundamental limits of collective gradient sensing are affected by these assumptions, we set up numerical stochastic and spatially-extended simulations of the system. Organoids were simulated using the HSim rule-based modeling program [16], version released 4/27/2015. For parameter exploration, a Python script generated model files with appropriate parameters and called HSIM with random seeds. Simulations were run on IBM NeXtScale nodes with Intel Xeon E5-2660 V2 and V3 processors.

Simulations were run for model organoids represented as coupled linear chains with the following numbers of cells: 3, 6, 10, 12, 15, 20, 25, and 50. For each simulated cell $n$, a set of molecules $(S_n, A_n, I_n, R_n)$ was initiated which interacted only with each other (Table S1). In the LEGI model, $S_n$ (the signal molecule) activates $A_n$ and $I_n$. The activated $A_n$ was allowed to activate $R_n$, and the activated $I_n$ was allowed to deactivate it. $I_n$ was also allowed to diffuse to become $I_{n\pm1}$. Each interaction was modeled as a Michaelis-Menten reaction. $A_n$ and $I_n$ were both allowed to deactivate with equal rates. Spherical cells with diameter 10 micron were initialized with $A_n = 1000$, $I_n = 1000$, $R_n = 500$, and $R_n^* = 500$ molecules. $S_N$ was initialized to 1000 in each simulation's final cell, with the gradient of 5 molecules per cell. All kinetic parameters present in both the theory and the simulations were selected to match (see Fig. 4 of the main text and Table S1). To investigate the effects of saturation, deactivation rates of $A_n$ and $I_n$ were scaled by 1/4 and 1/10 for partial and full saturation, respectively. High saturation of $A_n$ and $I_n$ was confirmed by removing reactions with $R_n$ and observing nonlinear response to varying $S_n$. The diffusion rate of $I_n$ was scaled accordingly to maintain a communication strength $n_0 = \sqrt{\gamma/\mu} = 10$ cells. Supplementary Table S1 shows the values of all kinetic rates used in the low saturation simulations.

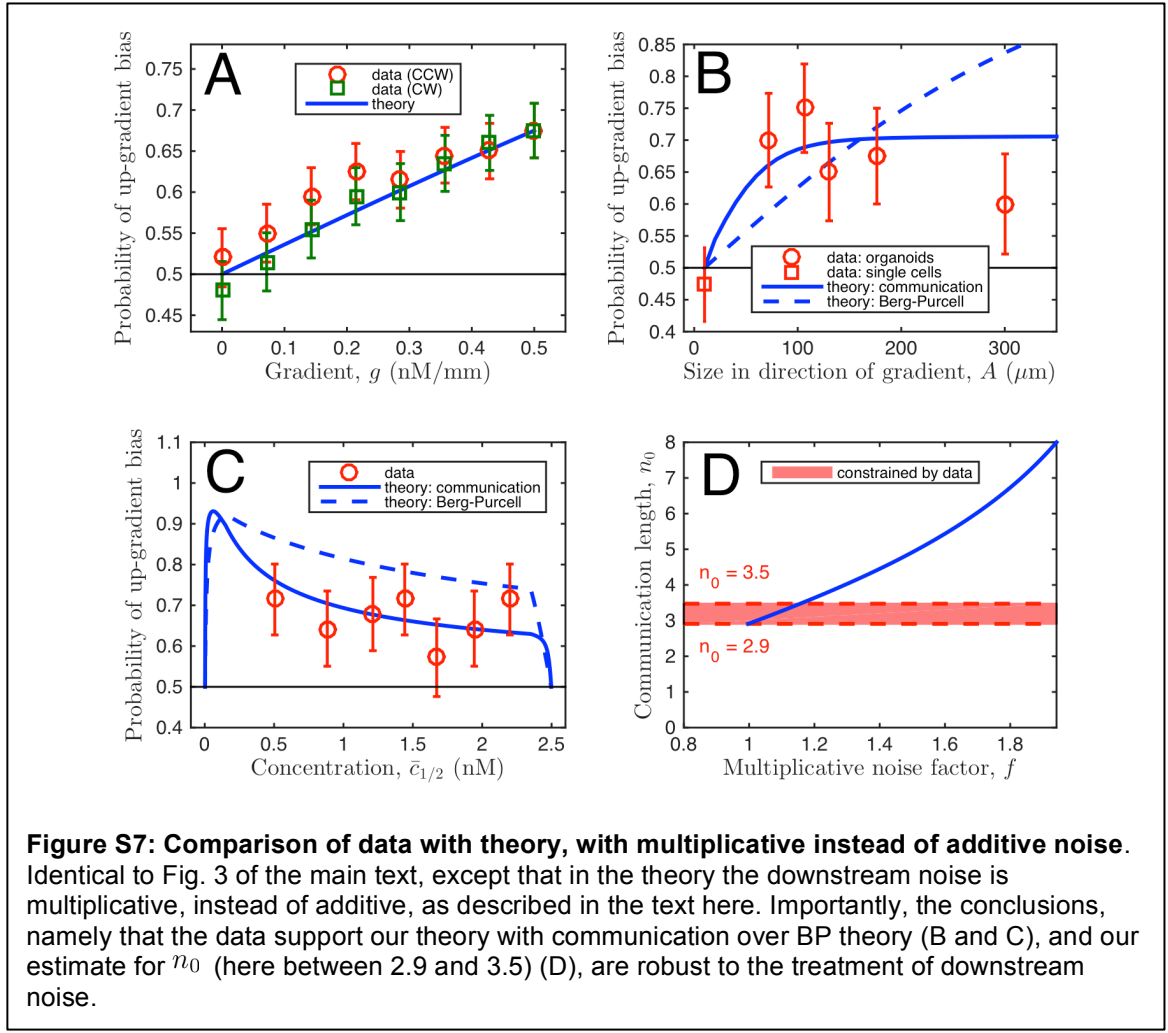

**Figure S7: Comparison of data with theory, with multiplicative instead of additive noise.** Identical to Fig. 3 of the main text, except that in the theory the downstream noise is multiplicative, instead of additive, as described in the text here. Importantly, the conclusions, namely that the data support our theory with communication over BP theory (B and C), and our estimate for $n_0$ (here between 2.9 and 3.5) (D), are robust to the treatment of downstream noise.

For each scenario (low, medium, and high saturation) and each number of cells, 16,384 simulations were run for a total of 393,216 runs. Simulations were ran sufficiently long (10,000 sec) so that the SNR had reached the steady state. SNR is reported as the squared mean over the variance of

$$\frac{R_n^* - R_n}{(R_n^* + R_n)/2}$$

in the final cell at the end of simulations. Error bars are determined by bootstrap sampling, reporting variance of 100 re-samples of size 16,384 taken from the original data with replacement.

**10. Limits on the size of the multicellular sensory unit with multiplicative downstream noise**

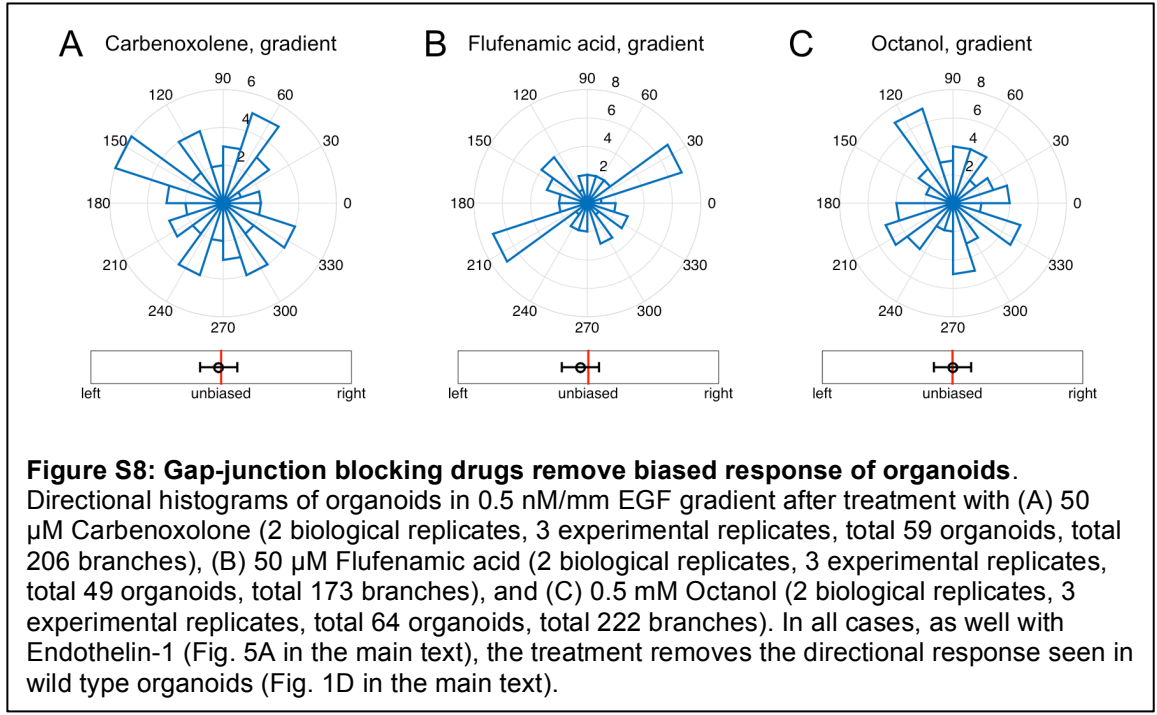

**Figure S8: Gap-junction blocking drugs remove biased response of organoids.** Directional histograms of organoids in 0.5 nM/mm EGF gradient after treatment with (A) 50 µM Carbenoxolone (2 biological replicates, 3 experimental replicates, total 59 organoids, total 206 branches), (B) 50 µM Flufenamic acid (2 biological replicates, 3 experimental replicates, total 49 organoids, total 173 branches), and (C) 0.5 mM Octanol (2 biological replicates, 3 experimental replicates, total 64 organoids, total 222 branches). In all cases, as well with Endothelin-1 (Fig. 5A in the main text), the treatment removes the directional response seen in wild type organoids (Fig. 1D in the main text).

In the main text, Fig. 3, we compared theoretical predictions of the BP model, as well as the model accounting for the communication noise, with the experimental data under the assumption that the noise in initiation of the phenotypic response, downstream of the gradient sensing, is additive. Here we consider a multiplicative noise model. For the BP theory, we again calculate the probability that the measured number of ligand molecules in the $N$'th cell is larger than in the first, $\nu_N > \nu_1$. However, now we take $\nu_n$ as Gaussian-distributed with mean $\bar{c}_n a^3$ and variance $\bar{c}_n a^3 f^2$, where $f^2 \geq 1$ represents the multiplicative increase due to downstream noise. Similarly, for our theory with diffusive communication, we calculate the probability that $\Delta_N > \Delta_1$, where $\Delta_n$ is Gaussian-distributed with mean $\bar{\Delta}_n$, and variance $(\delta \Delta_n)^2 f^2$, where both $\bar{\Delta}_n$ and $(\delta \Delta_n)^2$ are calculated earlier in this Supplementary Information. Supplementary Figure S7 is the multiplicative noise analog of Fig. 3 in the main text. Importantly, Fig. S7 demonstrates that our results depend only weakly on the assumed properties of the downstream noise. In particular, with either additive or multiplicative noise, the data support our theory with communication over BP theory (Fig. 3B and C of the main text, and Fig. S7B and C here), and we obtain similar estimates of the multicellular sensory unit given by $n_0$ (Fig. 3D of the main text and Fig. S7D here).

## 11. Treatments with gap junction-blocking drugs remove organoid response to EGF gradients

In addition to Endothelin-1, Fig. S8 confirms that other other gap-junction blocking drugs also remove the directional response of the organoids.

## 12. Calcium signaling is coordinated in nearby cells

To test the hypothesis that the global, diffusive inhibitory messenger in the organoids is related to calcium signaling (such as IP3 or calcium itself) we manually tracked 5 cells in the area at the front of a growing branch (see Supp. Movie 1) in an organoid derived from a transgenic mouse expressing genetically encoded Ca reporter GCaMP4, under the control of the CAG promoter [17], see Fig. S9. Calcium spikes in these cells are highly synchronized, indicating communication by calcium spikes inducing messengers. Note also that the size of the tip is consistent with our estimate of the gradient sensing unit (about 4 cells).

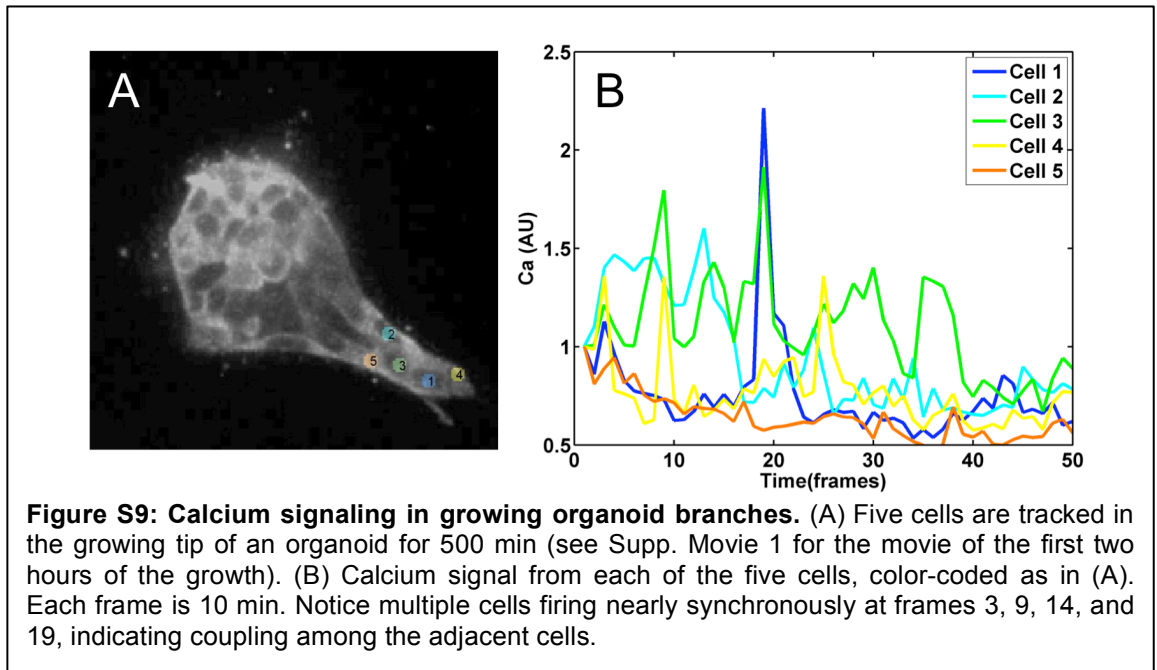

**Figure S9: Calcium signaling in growing organoid branches.** (A) Five cells are tracked in the growing tip of an organoid for 500 min (see Supp. Movie 1 for the movie of the first two hours of the growth). (B) Calcium signal from each of the five cells, color-coded as in (A). Each frame is 10 min. Notice multiple cells firing nearly synchronously at frames 3, 9, 14, and 19, indicating coupling among the adjacent cells.

### 13. Gradient establishment in the device

Numerical simulations show that a linear gradient of EGF, a 6.4 kDa protein, is established in our device in less than 24 hrs. We verify this by flowing an easily observable 10 kDa fluorescent protein (Dextran, Cascade Blue, Life Technologies) through the system and imaging it a day after the initition of the experiment. Supplementary Fig. S10, indeed, shows a nearly linear gradient. EGF is smaller, has a higher diffusion coefficient, and will establish a stable gradient even faster.

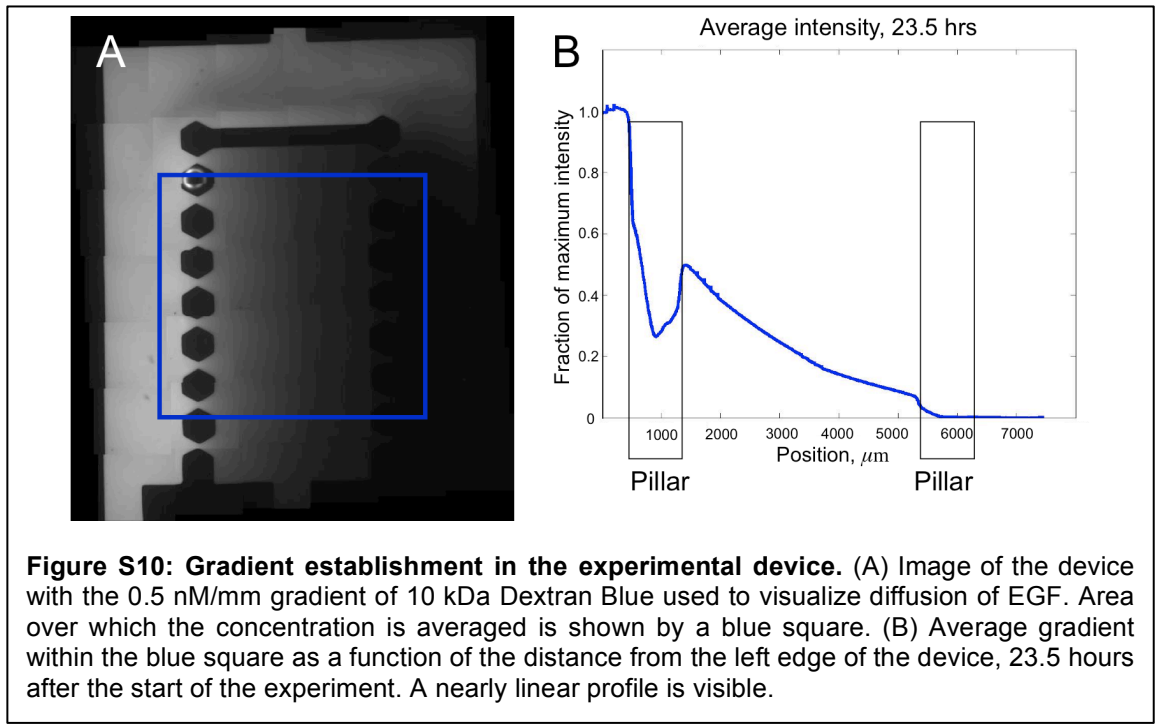

**Figure S10: Gradient establishment in the experimental device.** (A) Image of the device with the 0.5 nM/mm gradient of 10 kDa Dextran Blue used to visualize diffusion of EGF. Area over which the concentration is averaged is shown by a blue square. (B) Average gradient within the blue square as a function of the distance from the left edge of the device, 23.5 hours after the start of the experiment. A nearly linear profile is visible.

**References**

1. Berg, H.C. and E.M. Purcell, *Physics of chemoreception.* Biophys J, 1977. **20**(2): p. 193-219.
2. Endres, R.G. and N.S. Wingreen, *Accuracy of direct gradient sensing by single cells.* Proc Natl Acad Sci U S A, 2008. **105**(41): p. 15749-54.
3. Goodhill, G.J. and J.S. Urbach, *Theoretical analysis of gradient detection by growth cones.* J Neurobiol, 1999. **41**(2): p. 230-41.
4. Hu, B., et al., *Physical limits on cellular sensing of spatial gradients.* Phys Rev Lett, 2010. **105**(4): p. 048104.
5. Levchenko, A. and P.A. Iglesias, *Models of eukaryotic gradient sensing: application to chemotaxis of amoebae and neutrophils.* Biophys J, 2002. **82**(1 Pt 1): p. 50-63.
6. Lin, B., et al., *Interplay between chemotaxis and contact inhibition of locomotion determines exploratory cell migration.* Nat Commun, 2015. **6**: p. 6619.
7. Mugler, A., A. Levchenko, and I. Nemenman, *Limits to the precision of gradient sensing with spatial communication and temporal integration.* arXiv, 2015: p. 1505.04346
8. Thorne, R.G., S. Hrabetova, and C. Nicholson, *Diffusion of epidermal growth factor in rat brain extracellular space measured by integrative optical imaging.* J Neurophysiol, 2004. **92**(6): p. 3471-81.
9. Detwiler, P.B., et al., *Engineering aspects of enzymatic signal transduction: photoreceptors in the retina.* Biophys J, 2000. **79**(6): p. 2801-17.
10. Gadgil, C., C.H. Lee, and H.G. Othmer, *A stochastic analysis of first-order reaction networks.* Bull Math Biol, 2005. **67**(5): p. 901-46.
11. Usmani, R.A., *Inversion of a Tridiagonal Jacobi Matrix.* Linear Algebra and Its Applications, 1994. **212**: p. 413-414.
12. Usmani, R.A., *Inversion of Jacobis Tridiagonal Matrix.* Computers & Mathematics with Applications, 1994. **27**(8): p. 59-66.
13. Wartlick, O., A. Kicheva, and M. Gonzalez-Gaitan, *Morphogen gradient formation.* Cold Spring Harb Perspect Biol, 2009. **1**(3): p. a001255.
14. Iglesias, P.A. and A. Levchenko, *Modeling the cell's guidance system.* Sci STKE, 2002. **2002**(148): p. RE12.

15. Jilkine, A. and L. Edelstein-Keshet, *A comparison of mathematical models for polarization of single eukaryotic cells in response to guided cues.* PLoS Comput Biol, 2011. **7**(4): p. e1001121.
16. Amar, P., et al., *A stochastic automaton shows how enzyme assemblies may contribute to metabolic efficiency.* BMC Syst Biol, 2008. **2**: p. 27.
17. Zariwala, H.A., et al., *A Cre-dependent GCaMP3 reporter mouse for neuronal imaging in vivo.* J Neurosci, 2012. **32**(9): p. 3131-41.